# Is There an Earth-like Planet in the Distant Kuiper Belt?

Patryk Sofia Lykawka[1], Takashi Ito (伊藤孝士)[2,3]

[1] Kindai University, Shinkamikosaka 228-3, Higashiosaka, Osaka, 577-0813, Japan; patryksan@gmail.com

[2] Center for Computational Astrophysics, National Astronomical Observatory of Japan, Osawa 2-21-1, Mitaka, Tokyo 181-8588, Japan

[3] Planetary Exploration Research Center, Chiba Institute of Technology, 2-17-1 Tsudanuma, Narashino, 275-0016, Chiba, Japan

## Abstract

The orbits of trans-Neptunian objects (TNOs) can indicate the existence of an undiscovered planet in the outer solar system. Here, we used N-body computer simulations to investigate the effects of a hypothetical Kuiper Belt planet (KBP) on the orbital structure of TNOs in the distant Kuiper Belt beyond ~50 au. We used observations to constrain model results, including the well-characterized Outer Solar System Origins Survey (OSSOS). We determined that an Earth-like planet ($m$ ~ 1.5–3 $M_{\oplus}$) located on a distant (semimajor axis $a$ ~ 250–500 au, perihelion $q$ ~ 200 au) and inclined ($i$ ~ 30 deg) orbit can explain three fundamental properties of the distant Kuiper Belt: a prominent population of TNOs with orbits beyond Neptune's gravitational influence (i.e., detached objects with $q$ > 40 au), a significant population of high-$i$ objects ($i$ > 45 deg), and the existence of some extreme objects with peculiar orbits (e.g., Sedna). Furthermore, the proposed KBP is compatible with the existence of identified Gyr-stable TNOs in the 2:1, 5:2, 3:1, 4:1, 5:1, and 6:1 Neptunian mean motion resonances. These stable populations are often neglected in other studies. We predict the existence of an Earth-like planet and several TNOs on peculiar orbits in the outer solar system, which can serve as observationally testable signatures of the putative planet's perturbations.

## 1 Introduction

A trans-Neptunian object (TNO) is a member of the Kuiper Belt located beyond Neptune with a semimajor axis, $a > 30$ au. These rock and ice bodies are the remnants of planet formation in the outer solar system (Lykawka 2012; Prialnik et al. 2020). TNOs can reveal important information about the formation and dynamical evolution of the giant planets, such as their migration behavior, and the fundamental properties of the protoplanetary disk from which they originated (Nesvorny 2018; Gladman & Volk 2021). More than 1000 TNOs that belong to distinct dynamical classes have been discovered, allowing for important advances in our understanding of the trans-Neptunian region (Lykawka & Mukai 2007b; Gladman et al. 2008; Bannister et al. 2018; Bernardinelli et al. 2022). However, a single evolutionary model that explains the entire TNO orbital structure has not been developed (Gladman & Volk 2021). Here, we focus on the distant Kuiper Belt with $a$ = 50–1500 au and discuss four main constraints that a successful model must explain for that region.

1) A prominent population of *detached TNOs* with orbits beyond Neptune's gravitational influence and not locked stably into Neptunian mean motion resonance (MMR). Typically, detached TNOs with perihelia $q > 40$ au experience Gyr-stable and non-scattering orbits within the scattered disk reservoir[1] of the Kuiper Belt (Gladman et al. 2002; Lykawka & Mukai 2007b; Lawler et al. 2017). We used the orbital information on the AstDys observational database[2] to determine that the apparent fraction of observed detached TNOs to the entire TNO population at $a > 50$ au is ~10%. However, the intrinsic fraction must be several times larger, as observational biases work against discovering more distant or large-$q$ objects (Gladman et al. 2002; Petit et al. 2011; Shankman et al. 2017a; Kavelaars et al. 2020). Although estimations of the detached population can vary widely, they could be comparable to or larger than that of TNOs on orbits experiencing gravitational scattering by Neptune ($q$ = 25–40 au) (Gladman & Volk 2021). Here, we assumed that the intrinsic ratio of scattering and detached populations is ≤1 in the distant Kuiper Belt[3]. Oort cloud formation models, including galactic tides and passing stars, could generate only detached TNOs with at least $a$ greater than a few hundred of au (Morbidelli & Levison 2004; Kaib et al. 2011; Brasser & Schwamb 2015). However, these scenarios are currently disfavored regarding Kuiper Belt formation, thus they are excluded from our main discussions below (see Gladman & Volk 2021 for more details). As considered in several representative models (including the favored Neptune grainy migration scenario), scattering TNOs experiencing MMR with coupled Kozai interactions[4] or chaotic diffusion cannot explain the detached population, in particular members with lower orbital inclinations, $i < 20$ deg, or with $a > 150$–250 au (Gomes et al. 2005; Lykawka & Mukai 2006; Lykawka & Mukai 2007c; Gomes et al. 2008; Lykawka & Mukai 2008; Brasser & Morbidelli 2013; Sheppard et al. 2016; Nesvorny et al. 2016; Kaib & Sheppard 2016; Pike & Lawler 2017; Gladman & Volk 2021; Chen et al. 2022). On the other hand, in the rogue planet scenario, the perturbations of a short-lived planet with 1–2 Earth masses ($M_\oplus$) during the early solar system could have produced a substantial detached population at $a > 50$–100 au (Gladman & Chan 2006; Huang et al. 2022). Here, we focus on producing the entire detached population with an appropriate intrinsic proportion to the scattering one, as assumed above.

---

[1] Detached trans-Neptunian objects (TNOs) were likely scattered by Neptune in the past before acquiring their current detached orbits. We consider detached TNO objects with $q > 40$ au, because we are interested in objects strongly decoupled from gravitational interactions with Neptune over Gyr-timescales (Lykawka & Mukai 2007b; Malhotra 2019; Batygin et al. 2021). Thus, these objects are dynamically fossilized in the Kuiper Belt (Gladman & Volk 2021). Note that our definitions of detached ($q > 40$ au) and scattering ($q$ = 25–40 au) TNOs differ from the nomenclature presented in Gladman et al. (2008). Nevertheless, we verified that our results do not depend on this or similar classification schemes.

[2] https://newton.spacedys.com/astdys/

[3] During the final stages of the revision of this manuscript, we became aware of Beaudoin et al. (2023). That work estimated an intrinsic ratio of scattering to detached populations of 0.64 (with at least a factor of 2 error) within $a$ = 48–250 au for absolute magnitudes < 9. Beaudoin et al. (2023) used Gladman et al. (2008)'s nomenclature, thus making a comparison with our results less straightforward. In addition, that work did not constrain the $a > 250$ au or the $q > 50$ au regions. Given these uncertainties, we consider our intrinsic ratio assumption compatible with that estimate and valid for this work.

2) A statistically significant population of TNOs with high orbital inclinations ($i > 45$ deg; henceforth '*high-i TNOs*') not predicted by the Neptune migration, rogue planet, and several other outer solar system scenarios (Lykawka & Mukai 2007b; Gomes et al. 2008; Lykawka & Mukai 2008; Nesvorny & Vokrouhlicky 2016; Nesvorny et al. 2016; Kaib & Sheppard 2016; Lawler et al. 2017; Pike & Lawler 2017; Kaib et al. 2019; Nesvorny et al. 2020; Gladman & Volk 2021). The apparent fraction of high-$i$ TNOs within the observed population beyond 50 au is ~2%. These fractions become ~1% and ~6% for the scattering and detached populations, respectively. Given the strong observational biases against discovering high-$i$ TNOs (e.g., most surveys focused on small or near-ecliptic latitudes), the intrinsic population must represent a much more significant fraction (Petit et al. 2017; Gladman & Volk 2021). Here, we focus on producing the high-$i$ TNO population in the distant Kuiper Belt by conservatively assuming that the intrinsic high-$i$ fraction is at least 2%. We also assume this fraction should be at least 1% and 6% within the scattering and detached populations, respectively. In startling contrast, however, the abovementioned representative models produce populations confined at $i < 45$–50 deg, with only an intrinsic ~1% belonging to the high-$i$ class. These results are in line with Kaib et al. (2019), who claimed that models with external perturbations should be invoked to explain the high-$i$ population. Similarly, Gladman & Volk (2021) discussed scenarios to explain the existence of high-$i$ TNOs. Another intriguing feature is that scattering TNOs with $i > 50$ deg have been discovered in the distant Kuiper Belt. It is unlikely that MMR + Kozai dynamics could generate these orbits, because this mechanism predicts correlated high-$i$ and low-$e$ (large-$q$) detached-like orbits (Gomes 2003; Gallardo 2006a; Kaib & Sheppard 2016; Nesvorny et al. 2016; Sheppard et al. 2016). Furthermore, the existence of some unstable TNOs with $q = 15$–25 au (i.e., dynamically controlled by Uranus and Neptune) that possess $i = 47$–169 deg implies that the inclination distribution tail could extend to ~170 deg if high-$i$ TNOs are the main source reservoir of those objects (Table A1 in the Appendix). A high-$i$ reservoir in the distant Kuiper Belt has been suggested in the literature (Gladman & Volk 2021; Kaib & Volk 2023). On the other hand, because other mechanisms could generate high-$i$ objects with $q < 25$ au (Brasser et al. 2012; Gladman & Volk 2021), potential sources remain under debate

[4] We are aware that the term 'Kozai interaction' is becoming obsolete and should be replaced with the von Zeipel–Lidov–Kozai (or vZLK) interaction (e.g. Ito & Ohtsuka 2019; Tremaine 2023). Nevertheless, we use this term throughout this paper for backward compatibility as well as for simplicity.

(Kaib & Volk 2023). Summarizing, in addition to exploring the origin of high-$i$ TNOs beyond 50 au ($q$ > 25 au), we also investigated the possibility of a high-$i$ source reservoir in the distant Kuiper Belt region.

3) A subpopulation of *extreme TNOs* with peculiar orbits difficult to obtain using representative models, as discussed above. Extreme TNOs offer unique opportunities to better understand the outer solar system. For example, the discovery of (90377) Sedna and other TNOs with very large perihelia (>60 au) revealed the necessity to consider additional perturbations other than the four giant planets in the trans-Neptunian region (Brown et al. 2004; Trujillo & Sheppard 2014; Sheppard et al. 2019). Here, we define objects with $q$ > 60 au, $i$ > 60 deg, $q$ > 50 au ($i$ < 20 deg) and $i$ > 50 deg ($q$ < 40 au) as extreme TNOs. In addition to Sedna, we identify nine extreme TNOs (Table 1). For specific assumptions during Oort cloud formation (e.g., dense stellar environments), passing stars may have produced TNOs on Sedna-like orbits (Kaib et al. 2011; Brasser & Schwamb 2015; Kaib & Volk 2023). However, this scenario suffers from severe timing issues, among other concerns (Gladman & Volk 2021). Alternatively, Sedna-like TNOs may be produced during dynamical interactions with a rogue planet (Gladman & Chan 2006). However, the short-lived rogue planet's perturbations imply the low efficiency of this process. More importantly, it is unclear if all these models can reproduce the diversity of extreme TNOs as evinced here. For example, 2014 $YX_{91}$ is an extreme TNO with a very high inclination (62.9 deg) and on a scattering orbit ($q$ = 34.73 au). Such a high-$i$ and low-$q$ (high-$e$) orbit is unlikely to be produced by Kozai interactions during an MMR capture for objects interacting with Neptune, because it instead produces high-$i$/low-$e$ detached-like orbits via conservation of the quantity $\cos(i)\sqrt{1-e^2}$ (Gomes et al. 2005; Gallardo 2006a). This behavior has been demonstrated in several investigations of Neptune migration and dynamics of individual TNOs (Gomes et al. 2008; Brasil et al. 2014; Nesvorny 2015a, Nesvorny & Vokrouhlick 2016; Kaib & Sheppard 2016; Sheppard et al. 2016; Pike & Lawler 2017). The very few objects with $i$ > 50 deg obtained in these models are outliers that belong to the detached population. However, the small perihelion of 2014 $YX_{91}$ implies that this object is currently scattering. Furthermore, the inclinations of resonant dropouts and scattering populations obtained in representative models strongly concentrate within ~30–40 deg (Brasil et al. 2014; Kaib & Sheppard 2016; Nesvorny & Vokrouhlicky 2016; Nesvorny et al. 2016; Pike & Lawler 2017). Therefore, considering these reasons, a hypothetical MMR capture coupled with Kozai interactions and subsequent resonance dropout is not a viable explanation for the peculiar orbit of 2014 $YX_{91}$ ([5]). Similarly, the orbits of other extreme TNOs are difficult to explain via resonant interactions during or after Neptune's migration. Therefore, a comprehensive model should explain the properties of a parent population (detached or high-$i$) *and* the orbits of their extreme members.

4) A population of *Gyr-stable resonant TNOs* in various Neptunian MMRs. This work defines objects that exhibit resonant libration for timescales longer than 1 Gyr as stable resonant TNOs. Conversely, the transient resonant populations established via resonance sticking are not an appropriate diagnostic of long-term dynamical stability, as they are scattering or detached objects temporarily captured in MMR. As demonstrated elsewhere (Lykawka & Mukai 2007a; Lykawka & Mukai 2008) and in this study, stable resonant TNOs have been found at 2:1 ($a_{res}$ = 47.8 au), 5:2 ($a_{res}$ = 55.4 au), 3:1 ($a_{res}$ = 62.6 au), 4:1 ($a_{res}$ = 75.8 au), 5:1 ($a_{res}$ = 88.0 au), and 6:1 ($a_{res}$ = 99.4 au), among other MMRs within 100 au. Most of these stable resonant TNOs remain locked in MMR for 4 Gyr, which suggests a primordial origin during the early solar system. There is insufficient discussion on the origin and preservation of these stable resonant populations within the framework of the representative Kuiper Belt models.

[5] To reproduce the current orbit of 2014 $YX_{91}$ via Kozai oscillations in such a hypothetical MMR capture (e.g., 5:2) during or after Neptune's migration, the initial inclination must be $i$ > 57–59 deg. Thus, the very high $i$ cannot be explained by this mechanism.

A successful model for the distant Kuiper Belt should explain these constraints simultaneously. The model should also avoid overpopulating the orbital space at $a < 100$ au with detached objects on small-to-moderate inclinations (e.g., $i < 30$ deg) because, although the probability of discovery is higher at small distances, observations have not confirmed a preference for lower-$i$ detached TNOs in this region (Bannister et al. 2018). The representative models discussed above and other outer-solar-system models (Hahn & Malhotra 2005; Nesvorny 2011; Batygin et al. 2012; Nesvorny 2015b; Nesvorny 2018; Clement et al. 2021) considering four to six giant planet instability/migration evolutions have yet to demonstrate that they could satisfy *all* of the main constraints consistently[6]. In conclusion, the distant Kuiper Belt probably did not form solely due to perturbation by the current four giant planets or hypothetical giant planets (or rogue planets) during/after the giant planets' migration.

It is likely that a few thousand dwarf planets (objects with a mass comparable to Pluto or Eris) and several tens of sub-Earth and Earth-class planets formed in the outer solar system during planet formation (Nesvorny & Vokrouhlicky 2016; Shannon & Dawson 2018; Nesvorny 2018). These bodies were mostly lost through gravitational scattering or collision with planets during their formation. Indeed, some models have considered the effects of rogue planets with masses comparable to Mars or Earth in the early solar system (Gladman & Chan 2006; Silsbee & Tremaine 2018; Huang et al. 2022). However, these studies did not investigate the formation of the Kuiper Belt; thus, they only demonstrated that perturbations of rogue planets could lead to detached TNO orbits. On the other hand, the existence of a trans-Neptunian resident planet has been used to explain distinct properties of the Kuiper Belt (Lykawka & Mukai 2008; Trujillo & Sheppard 2014; Malhotra 2018). Trujillo & Sheppard (2014) proposed that a 2–15 $M_\oplus$ planet located at 200–300 au is necessary to explain the alleged orbital angular alignments of TNOs located beyond 150–250 au (Trujillo & Sheppard 2014; Sheppard & Trujillo 2016). This proposal led to the Planet 9 model, which favors a 5–10 $M_\oplus$ super-Earth planet on a distant orbit beyond ~400 au with $q > 300$ au (de la Fuente Marcos & de la Fuente Marcos, 2014; Batygin & Brown 2016; de la Fuente Marcos & de la Fuente Marcos 2017; Batygin et al. 2019; Brown & Batygin 2021). However, this model was not designed to explain the Kuiper Belt's orbital structure, and many studies disfavor the validity of the alleged TNOs' angular alignments (Lawler et al. 2017; Shankman et al. 2017a; Shankman et al. 2017b; Gladman & Volk 2021; Napier et al. 2021; Bernardinelli et al. 2022). The super-Earth proposed by the Planet 9 model should not be confused with the sub-Earth-to-Earth-class hypothetical planets proposed in the literature (Lykawka & Mukai 2008 and references therein) (e.g., 'Planet X') and this study. Indeed, Planet 9 is much more massive and hypothesized to be located on more distant orbits. Also, while Planet 9 was proposed to address possible peculiar properties of some distant TNOs, our scenario addresses the structure of the distant Kuiper Belt by considering the abovementioned constraints.

[6] The Nice model (Nesvorny 2018 and references therein) is currently the leading giant planet instability model. However, on close inspection of the orbital structure of their obtained Kuiper Belts, this model apparently cannot explain the main constraints of the distant Kuiper Belt. Other shortcomings of the Nice model have been discussed elsewhere (Gladman et al. 2012; Pike et al. 2017; Lawler et al. 2018a; Lawler et al. 2019; Volk & Malhotra 2019).

Here, we consider the influence of Mars- and Earth-class planets on the formation of the distant Kuiper Belt. We refer to these planets as Kuiper Belt planets (KBPs). Lykawka & Mukai (2008) proposed that a sub-Earth KBP (0.3–0.7 $M_{\oplus}$) located at 100–250 au with $q$ > 80 au and $i$ = 20–40 deg was required to explain the orbital structures and other properties of the entire trans-Neptunian region. Volk & Malhotra (2017) found that a KBP with mass of up to 2.4 $M_{\oplus}$, located within 100 au, could explain the plane of the Kuiper Belt (but see van Laerhoven et al. 2019 for a different conclusion). Early solar-system interactions between the giant planets and several Mars- and Earth-class objects may have produced a resident planet within 200 au with $q$ = 40–70 au and $i$ < 30 deg (Silsbee & Tremaine 2018). Lykawka & Mukai (2008) demonstrated that a KBP could produce detached, high-$i$, and transient resonant TNOs (via resonance sticking), but at the same time effectively destroy the stability of Gyr-stable resonant TNOs, depending on the KBP's mass and orbit. A stable resonant TNO may evolve for >1 Gyr timescales protected against close gravitational encounters with Neptune. However, the gravitational perturbations of a KBP can disrupt this stability, so resonant escapees may acquire scattering or detached orbits. Because stable populations are well established in several resonances, the stability of the high-eccentricity resonant members can constrain the perihelion of a KBP. Following this premise, Lykawka & Mukai (2008) found that any KBP must have $q$ > 80 au to preserve the stable populations in the 3:2, 2:1, and 5:2 MMRs identified at that time. The results of Lykawka & Mukai (2008) and this study rule out the existence of many hypothetical planets, including those proposed by Silsbee & Tremaine (2018). Similarly, observational constraints do not imply the presence of KBPs within $a$ ~ 100 au (Iorio 2017; van Laerhoven et al. 2019).

We used extensive simulations to demonstrate that a resident undiscovered Earth-like planet could explain the four main constraints in the distant Kuiper Belt. Furthermore, because a KBP's perturbations can shape orbital structure in the trans-Neptunian region over the age of the solar system, the orbital properties of the various known TNO populations can constrain the existence of a KBP (Lawler et al. 2017; Shannon & Dawson 2018; Nesvorny 2018; Lawler et al. 2018a). It is plausible that a primordial planetary body could survive in the distant Kuiper Belt as a KBP, as many such bodies existed in the early solar system.

## 2 Methods

We used N-body simulations to investigate the dynamical evolution of real and model TNOs in the distant Kuiper Belt. The four giant planets on their current orbits were considered in our control model throughout the investigations described herein. First, we verified long-term resonance occupancy within the strongest MMRs (n:1 or n:2) covering the region from 2:1 ($a_{res}$ ~ 48 au) to 40:1 MMR ($a_{res}$ ~ 352 au) using our Restick code (Gallardo 2006b; Lykawka & Mukai 2007a; Lykawka & Mukai 2007c; Yu et al. 2018) and visual inspection as needed. Orbital information for these objects was retrieved from the AstDys observational database in late February 2021. Specifically, in this investigation we considered the best-fit orbits with small orbital uncertainties (median $da/a$ = 0.025%). Stable resonant TNOs were identified by selecting objects that exhibited continuous resonant libration for at least 1 Gyr and up to 4 Gyr during their orbital evolution. As a complementary investigation, we also identified stable resonant objects among the intrinsic CFEPS L7 resonant populations in the 2:1, 5:2, 3:1, and 5:1 MMRs (Petit et al. 2011; Petit et al. 2017). A summary of the results is given in Table 2. Further details and implications of these results are discussed in Section 3.1.

Next, we selected Gyr-stable resonant TNOs and L7 objects and conducted new simulations up to 4 Gyr by including a KBP in the system. We repeated this experiment for a total of 140 KBP models after considering a wide range of putative planet orbits and masses. Then, we compared the resulting stable resonant populations in our five-planet systems (the four giant planets and a KBP) with those of the control system. Because a KBP can disrupt the stable resonant populations, we constrained the orbits and masses of various hypothetical KBPs by requiring these populations to retain their main properties at the end of the simulations. When comparing the final populations with those of the control model, we adopted the following success criteria for each resonance: at least 1/3 of the high-$e$ component ($q$ < 33 au) of the resonant population should occupy similar orbits, the median eccentricity of the resonant population should be compatible with that from observations or the L7 intrinsic population, and the resonant population should retain at least 1/4 of the original population. We note that the first criterion is the best diagnostic of resonant population disruption and is weakly dependent on observational biases, while the second and third offer useful complementary insights. Further details are provided in Tables A2 and A3 in the Appendix.

We progressed to the main stage of our investigation after the properties of KBPs that could preserve the known stable resonant populations had been determined. Although a KBP with appropriate orbit and mass would be compatible with the existence of stable resonant TNOs, the putative planet would strongly perturb the orbits of TNOs in the scattered disk. A key question is whether this perturbation could explain the observed distinct populations of TNOs in the distant Kuiper Belt. Conversely, these populations may provide observational signatures for the existence of a KBP in the outer solar system. To test this hypothesis, we investigated the long-term dynamical evolution of primordial scattered disks using 61 models of KBPs (Table 3). To the best of our knowledge, there is no consensus about the properties of the primordial scattered(ing) disk. The scattering population component obtained in simulations by Gladman & Chan (2006) originated the intrinsic CFEPS L7 scattering population (Petit et al. 2011; Petit et al. 2017). The model results of Kaib et al. (2011) were considered in several studies of the dynamical evolution of scattering populations (Shankman et al. 2013; Shankman et al. 2016; Lawler et al. 2017; Yu et al. 2018). Pike et al. (2017) also analyzed the scattering objects produced in a Nice model scenario. Although these studies focused on the dynamical evolution over the past 1 Gyr, they found that the orbital properties of an evolving scattering population do not depend on the specifics of the scattering events nor the models considered. Furthermore, we analyzed the orbital distributions ($a$, $i$, and $q$) of the scattering population component obtained during the first ~100–500 Myr of 4-Gyr-simulations by Kaib et al. (2011) (personal communication). Overall, we found that these distributions were similar to those of the L7 scattering population. Although further studies are needed to understand better the early scattered disk (e.g., Huang et al. 2022), we conclude that its main property was confinement in perihelia (<40 au). For these reasons, we used the L7 scattering population as representative of the primordial scattered disk (Petit et al. 2011; Petit et al. 2017) with an inclination correction as applied in previous studies (Kaib et al. 2011; Shankman et al. 2013; Alexandersen et al. 2013). The inclination followed the probability density distribution:

$$\mathrm{f}(i) \propto \sin(i) \exp^{-\frac{(i-<i>)^2}{2\sigma^2}} \quad (1)$$

where $<i>$ = 7 deg and $\sigma$ = 11.7 deg.

To obtain statistically meaningful results, we generated 29 clones of the L7 scattering population, in which the original semimajor axis and perihelion values were randomly scaled by ± 20% and ± 1%, respectively. This procedure resulted in 48,060 objects in our primordial scattered disk (Figure 1). We verified that our initial population's perihelia were well within 40 au (e.g., only 1% of the population was initialized with $38 < q < 40$ au). Also, 99.70% (99.93%) of this population was confined to $i < 45$ (50) deg initially. Finally, we evolved this population in the 61 and control models until 4.5 Gyr.

A modified version of the MERCURY integrator was used to conduct the simulations (Chambers 1999; Hahn & Malhotra 2005). The simulation time step was 0.6 yr to ensure reliable calculations in the outer solar system. The effects of the galactic tides and passing stars were not included in our modeling, as they may play a role only at $a$ > 1000–2000 au (Gladman et al. 2008; Lawler et al. 2017; Nesvorny et al. 2017; Silsbee & Tremaine 2018; Kaib et al. 2019). Besides, we were interested in investigating the sole perturbation effects of a KBP. Objects that evolved to heliocentric distances less than 1 au or greater than 2500 au (approximately $a$ > 1250 au for typical scattering objects) were discarded from the simulations.

After obtaining systems that represented the distant Kuiper Belt at the end of 4.5 Gyr, we used the Outer Solar System Origins Survey (OSSOS) Survey Simulator (OSS) to compare these results with observations as constrained by the well-characterized OSSOS+ surveys (Petit et al. 2011; Alexandersen et al. 2016; Bannister et al. 2016; Petit et al. 2017; Bannister et al. 2018). This procedure is possible because the OSS simulates detections based on a population provided by the user (e.g., model results), which are later compared with OSSOS observations. More OSS details are given by Lawler et al. (2018a). However, few detached and high-*i* TNOs were discovered by OSSOS+ surveys; thus, it is challenging to discriminate models that produce objects with such orbits in the far outer solar system (Lawler et al. 2017). For this reason, in this work a model was deemed successful only if it was compatible with observations after it had been debiased using the OSS *and* if it satisfied the main constraints discussed in Section 1.

## 3 Results and Discussion

### 3.1 Dynamical stability and other properties of Gyr-stable resonant populations

We identified resonant TNOs in the outer solar system that are stable over Gyr timescales in the 2:1, 5:2, 3:1, 4:1, 5:1, and 6:1 MMRs (Table 2). Several of these objects are likely primordial, as evidenced by the 4-Gyr-stable members. Although transient resonance sticking can produce 1-Gyr-timescale captures of TNOs in MMRs within 100–150 au (Lykawka & Mukai 2007c; Yu et al. 2018), these are low-probability events. Furthermore, our simulations revealed that many stable resonant TNOs have full-width libration amplitudes, *A*, that are much smaller than would be expected from resonance sticking. This finding was true for several 2:1 resonant TNOs with $A < 20$ deg in asymmetric libration and 5:2 resonant TNOs with $A < 100$ deg. Despite the relatively small-number statistics for the 3:1 and 4:1 MMRs, the inclination distributions of 4-Gyr-resident resonant TNOs also imply that a fraction of the stable 2:1 and 5:2 resonant population origins differ from those of their counterparts in the 3:1 and 4:1 MMRs. For the populations with 4-Gyr stability, the median inclinations of the 2:1 and 5:2 stable resonances were 7.5 and 11.3 deg, respectively, significantly smaller than the 23.2 deg and 26.5 deg of the two more distant MMRs mentioned above. Regarding perihelia, the medians were 35.3 and 33.0 au for the 2:1 and 5:2 MMRs, respectively, again smaller than 38.0 and 43.0 au for the 3:1 and 4:1 MMRs. The Kozai mechanism can operate in all these MMRs, so it probably cannot explain the differences between the 2:1–5:2 and 3:1–4:1 groups. Instead, these features may indicate that a significant population component was captured via resonance sweeping from the protoplanetary disk during Neptune's outward migration (Malhotra 1995; Hahn & Malhotra 2005; Lykawka & Mukai 2007c; Pike & Lawler 2017). These findings are also consistent with the idea that the bulk populations in the 3:2, 2:1, and 5:2 MMRs did not originate via resonance sticking (Yu et al. 2018). Also, two stable resonant TNOs in the 3:1 (2016 $SO_{56}$) and 4:1 (2008 $UA_{332}$) resonances in asymmetric libration have $A \sim 10$ deg, which probably cannot be explained by resonant sticking. These results imply that Neptune captured stable resonant TNOs from the disk up to the 4:1 MMR with resonance sweeping. Because the stable resonant populations in the 3:1–4:1 group apparently lack the low-*i* concentration seen in the 2:1–5:2 group, the 3:1 and other distant MMRs probably swept a dynamically hot region located beyond ~44–58 au for a migrating Neptune located at 21 (Hahn & Malhotra 2005; Lykawka & Mukai 2007b) or 28 au (Nesvorny 2015a; Kaib & Sheppard 2016; Nesvorny & Vokrouhlicky 2016). This hot region could represent an extension of the protoplanetary disk perturbed to higher *e*, *i* (Lykawka & Mukai 2007c) or a component of an already stirred primordial scattered disk before planetary migration.

Irrespective of the origin of the stable resonant TNOs, the existence of these populations strongly constrains the orbits and masses of undiscovered KBPs. The high-*e* resonant members are most sensitive to perturbations of the putative planet because they spend more time at large distances, and thus are more prone to leave the resonance owing to close encounters with a KBP. Therefore, the aphelia of our identified high-*e* stable resonant TNOs imply that a KBP should have a perihelion larger than ~115 au to preserve stable resonant TNOs up to 4:1 MMR. The constraints are weaker for the 5:1 and 6:1 MMRs because of the small-number statistics (Table 2). Nevertheless, for completeness, we considered all identified resonant stable populations when evaluating the perturbation effects of a KBP (Section 2).

The effects of Mars-mass KBPs (0.1 $M_\oplus$) are modest for a wide range of orbits, and even KBPs with $q < 115$ au might yield acceptable outcomes. However, the more massive KBPs ($m \geq 0.3$ $M_\oplus$) tend to perturb stable resonant populations significantly, so KBPs with larger perihelia are necessary to meet the stable resonant TNO constraint. In summary, the results imply that a Mars-mass KBP with $a > 175$ (>250) au and $q > 95$ (>65) au would be consistent with this constraint. For sub-Earth KBPs, $a > 175$ au and $q > 120$ au ($m = 0.3$ $M_\oplus$) or $q > 145$ au ($m = 0.5$ $M_\oplus$) are required. The requirements for more massive Earth-class KBPs (1–3 $M_\oplus$) are $a > 250$ au and $q > 145$ au ($m = 1$ $M_\oplus$) or $q > 195$ au ($m = 2$ or 3 $M_\oplus$).

The most distant stable resonant TNO identified is represented by a single object locked within the 6:1 MMR [(528381) 2008 $ST_{291}$]. If it is found that this resonance has a significant Gyr-stable population, the required perihelion for the KBP would be a few tens au greater than the above values. In addition to the 6:1 MMR, the possible existence of stable resonant populations in more distant resonances (7:1, 8:1, …, n:1) would further constrain the orbits/masses of hypothetical distant planets and rule out the main trans-Neptunian planets proposed thus far[7] (Lykawka & Mukai 2008; Batygin & Brown 2016; Brown & Batygin 2021). As a result, it is crucial that dedicated surveys characterize the orbits of new TNOs beyond 100 au with sufficient accuracy to allow for verification of their long-term stability in distant MMRs.

### 3.2 Dynamical evolution of the primordial scattered disk

First, the orbital distribution of our scattered disk obtained in the control model is remarkably similar to the results of representative models of the Kuiper Belt (Lykawka & Mukai 2007c; Gomes et al. 2008; Sheppard et al. 2016; Nesvorny et al. 2016; Kaib & Sheppard 2016; Pike et al. 2017). The most notable structure is the detached objects created within ~250 au and having $i > 20$ deg due to the dynamics of strong n:1 and n:2 MMRs + Kozai interactions (Figure 2). The control model also yielded a ratio of scattering and detached populations $r_{sca/det} = 4.18$ and a high-*i* fraction of 1.2% beyond 50 au (Table 3). Finally, none of the extreme TNOs were reproduced in this model.

[7] The 8:1 and 9:1 resonant TNOs reported in the literature (Volk et al. 2018; Bernardinelli et al. 2022) are likely members of the transient resonant populations, so they do not belong to the Gyr-stable resonant populations considered in our analysis. Consistent with the results of Volk et al. that resonance sticking was a plausible explanation for the observed 9:1 population, we found that our identified 9:1 resonant TNOs, as well as the 8:1 resonant TNOs, were not Gyr-stable. Finally, Crompvoets et al. (2022) concluded that the population ratios between various MMRs beyond 50 au were consistent with resonance sticking predictions.

Thus, our control model considering solely the four giant planets failed to explain the detached, high-$i$, and extreme TNO populations. This model also confirmed the main results reported by several previous studies, as discussed in Section 1.

For these reasons, we focus below on the results of models that included a KBP in the system.

3.2.1 Detached population

Our simulations revealed that more distant KBPs preserved the stable resonant populations and simultaneously strongly perturbed the primitive scattering populations. In this way, these simulations produced new populations of TNOs in the scattered disk that can serve as observational signatures for the existence of a KBP. The orbits of these objects correlate with the KBP's semimajor axis, inclination, and mass in orbital space, as detailed below.

We began our analysis with Mars-mass KBPs. These planets produced modest effects in the scattered disk. Although all KBPs tested with $m$ = 0.1 M$_\oplus$ produced some detached objects resulting in ratios of scattering and detached populations $r_{sca/det}$ = 1.89–3.38 (lower than 4.18 found for the control system), these ratios were higher than those estimated from observations in which $r_{sca/det}$ was ≤1 (Section 1 and Table 3). Furthermore, scattered disks perturbed by Mars-mass KBPs displayed inclination distributions similar to the control system, with inclinations confined to within 45 deg (~99%). This result implies that such KBPs cannot explain the high-$i$ TNOs discussed in Section 1. While 2012 VP$_{113}$- or 2013 SY$_{99}$-like objects may be produced depending on a KBP's orbit (albeit with low efficiency), no model considering 0.1 M$_\oplus$-KBPs produced analogs of other extreme TNOs. Therefore, a Mars-mass KBP with $a$ = 250–500 au, $i$ = 0–30 deg, and $q$ = 65–95 au cannot explain the main constraints in the distant Kuiper Belt. Considering that the effects of such planets would be too small to yield observational signatures compatible with observations, these results might apply to any Mars-mass KBP located beyond 65 au.

The effects of more massive KBPs ($m$ ≥ 0.3 M$_\oplus$) were critical in the scattered disk. Most of these KBP models produced prominent detached populations, resulting in ratios $r_{sca/det}$ = 0.49–2.09. In particular, ratios $r_{sca/det}$ ≤ 1 more consistent with observational constraints were obtained only for KBPs with $m$ ≥ 1 M$_\oplus$ (Section 1 and Table 3). Another significant achievement is that these models can easily produce low-$i$ detached TNOs, which were difficult to obtain in previous models. At the same time, only 5–10% of these objects reside at $a$ < 100 au, so our model avoided overpopulating this region, in agreement with additional observational constraints. Furthermore, models including KBPs with $m$ ≥ 1 M$_\oplus$ often produced analogs of the extreme TNOs with large perihelia (2012 VP$_{113}$, Sedna, and 2013 SY$_{99}$). Models considering more massive and more distant KBPs ($m$ ≥ 1.5 M$_\oplus$ and $a$ ≥ 350 au) yielded even some analogs of (541132) Leleakuhonua and 2021 RR$_{205}$ within the farthest region investigated in this work at $a$ ~ 1000–1250 au. However, this region suffers from small-number statistics due to computational limitations in our simulations (Section 2). A more detailed investigation of $a$ > 1000 au extreme TNOs is warranted.

In conclusion, a massive KBP located at $a$ = 250–500 au and $q$ > 195 au ($m$ = 1–3 M$_\oplus$) can simultaneously produce the detached population and the orbits of extreme TNOs. Representative cases corroborating these results are illustrated in Figures 3–7. The right panels of Figure 8 also illustrate the distributions of perihelia for some of our best KBP models (see Section 3.3 for more details).

3.2.2 High-$i$ population

We analyzed the inclination distributions of our scattered disks and focused on KBPs with $m \geq 1$ M⊕ that yielded more promising results, as discussed above. For KBPs with $m = 1$ M⊕, we found that only those on inclined orbits ($i = 30$ deg) could produce a high-$i$ population approximately 1.5–8 times larger than that found in the control model. Specifically, this population represented 1.9–9.6% of the population beyond 50 au (KBP models #24, #26, #28, and #30 in Table 3). However, an Earth-mass KBP with $i \leq 30$ deg might not provide sufficient perturbation to explain a large population with $i > 45$ deg within the scattering population. Only Earth-class KBPs with $m = 1.5$–3 M⊕ satisfied the minimum 1% high-$i$ fraction among the scattering population, with an apparent preference for more eccentric ($q = 195$ au) rather than lower-$e$ ($q = 245$ au) orbits. Similarly, some massive KBPs yielded a population of objects with $i > 90$ deg that could be a source of TNOs and comets on retrograde orbits in the solar system (such as Halley-type), as shown in Figures 5–7. This result supports the hypothesis that a high-$i$ reservoir in the distant trans-Neptunian region could be sourcing the solar system with high-$i$ objects, including retrogrades (Gladman et al. 2009; Gladman & Volk 2021; Kaib & Volk 2023). Based on the results of our favored KBPs shown in bold in Table 3, the intrinsic high-$i$ fraction of our populations located beyond 50 au varied between 5.5% and 26.3%. The high-$i$ fractions of the scattering and detached populations were approximately 1–2% (1–4%) and 9–28% (20–42%), respectively, for KBPs with $i = 1$–15 deg ($i = 30$ deg). These results imply that KBPs with higher inclinations can increase the production efficiency of high-$i$ TNOs. Similarly, the fractions of scattering and detached objects on retrograde orbits were 0–1% (0–2%) and 0–18% (1–23%) for the same KBP models. Overall, our favored KBP models simultaneously satisfied all of the high-$i$ population constraints (Table 3). Regarding analogs of the extreme high-$i$ TNO 2014 $YX_{91}$, objects that acquired similar orbits (i.e., $i = 50$–70 deg and $q < 40$ au) were less common but also identified at the end of the simulations. Considering that this TNO is dynamically unstable with $q = 34.73$ au, such objects may be continuously replenished from a high-$i$ reservoir similar to the one discussed above.

In conclusion, the perturbations of a massive KBP with $m = 1.5$–3 M⊕ on a mildly inclined and eccentric orbit establish a high-$i$ reservoir of scattering and detached populations that could explain the high-$i$ TNOs and continuously supply unstable TNOs with $q = 15$–25 au and $i > 45$ deg (Figures 3–7). In conjunction with the constrained existence of Gyr-stable resonant TNOs, these results imply that an Earth-like KBP with $a = 250$–500 au, $q = 195$–245 au, and $i = 30$ deg would be able to produce the detached population, high-$i$, and extreme TNOs consistently. Ten optimal KBP models out of the favored 20 indicated in Table 3 (in bold) satisfied these orbital constraints.

Models that consider the perturbations of the four giant planets, and the external forces of Galactic tides and passing stars, imply that inner-Oort-cloud objects may be the source of the high-$i$ population with $q > 10$ au (Kaib et al. 2019; Kaib & Volk 2023). However, these models assume large populations of inner-Oort-cloud objects, an assumption that is not widely recognized. Furthermore, these models were designed to explain the Oort cloud and the delivery of comets from that reservoir. Therefore, it is unclear whether such models can be used to explain other constraints related to the distant Kuiper Belt formation. However, as discussed previously, the KBP model may explain high-i TNOs and satisfy other Kuiper Belt constraints.

### 3.3 Comparisons of model results with observations

#### 3.3.1 Observational constraints from the OSSOS Survey Simulator (OSS)

Our optimal KBP models predict the existence of substantial populations of yet undiscovered TNOs with large-$q$ and/or high-$i$ at $a > 150$ au, as indicated by some representative cases illustrated in Figures 3–7. To address this point and further constrain the best candidate KBP in the distant Kuiper Belt, we compared our favored systems' orbital distributions at the end of simulations with observations using the OSS. First, we ran the OSS for each KBP model until 5,000 objects were detected without restricting the orbits. Then, we considered detected objects with $a > 50$ au and $q > 25$ au that span the core of the distant Kuiper Belt[8]. This restriction also made our comparison with OSSOS observations more reasonable because we did not consider the formation of the classical Kuiper Belt within 50 au nor Neptune's resonance sweeping of that region. In this way, typically 3,000 detections per model were compared to 125 TNOs characterized by OSSOS for the same orbital range considered. This process was computed 10 times using different OSS random seeds for each KBP model. Finally, the default underlying absolute magnitude $H$-distribution was the same as the preferred one by Lawler et al. (2018b), namely, a broken power-law distribution with a bright-end slope of 0.9, a faint-end slope of 0.5, a $H_{\mathrm{break}} = 8.3$ mag, and a divot of contrast 3.2. In the literature, it has been shown that the $H$-distribution in the scattered(ing) disk is better described by a divot or a knee with similar distribution properties, while a single slope is rejected with high confidence (Shankman et al. 2013; Shankman et al. 2016; Pike et al. 2017; Lawler et al. 2017; Lawler et al. 2018b; Kaib et al. 2019; Crompvoets et al. 2022). These works generally found similar results when comparing the divot and knee distributions. For completeness, we also tested the favored knee distribution in the literature (e.g., Lawler et al. 2018b) with a bright-end slope of 0.9, a faint-end slope of 0.4, and $H_{\mathrm{break}} = 7.7$ mag among our optimal KBP models. For these reasons, we believe that our choice of the $H$-distribution was appropriate and that its details should not affect the main results of this work.

The entire populations with $a > 50$ au, $q > 25$ au, and $i > 0$ deg were considered when making comparisons of model distributions with observations using the OSS, as reported below. In particular, the semimajor axis, inclination, and perihelion distributions of detected objects and

[8] We did not consider unstable objects with $q = 15$–$25$ au in this investigation because other mechanisms not modelled in this work (e.g., Galactic tides and passing stars) could generate objects with such orbits. We also removed the 4-Gyr-stable resonant TNOs identified in the 5:2 MMR from the observed sample, because this population likely formed by Neptune's resonance sweeping of the protoplanetary disk within ~50 au. Further justifications can be found in Section 3.1.

OSSOS observations for some of our best models are illustrated in Figure 8. A representative example comparing model detections with OSSOS observations in orbital space is also shown in Figure 9. This figure also highlights the strong observational biases against the discovery of TNOs on distant (large $a$ or large $q$) and high-$i$ orbits. Although there are only 3, 2, and 16 TNOs with $a >$ 250 au, $i >$ 45 deg, and $q >$ 40 au in our observational sample, respectively, we stress that the perturbations of our KBPs affected the detection distributions in much wider regions of $a$-$q$-$i$ space. In particular, this was the case for $a >$ 100 au (18 TNOs), $i >$ 15 deg (57 TNOs), and $q >$ 37 au (53 TNOs), as discussed below (see also Figure 8).

Overall, the distributions of detected model and real objects are similar; however, some caution is needed regarding their interpretation. While the inclination and perihelion distributions roughly match the OSSOS data, the semimajor axis distributions do not (Kolmogorov-Smirnov [K-S] test probability <0.2%). Although the KBP model notably improved the $a$-distribution compared to the control model, it predicts more (less) detected objects at large $a$ (small $a$) than observed. Curiously, we note that a similar semimajor axis mismatch plagues representative models that compared model scattered disk objects with observations (Lawler et al. 2017; Lawler et al. 2018b; Beaudoin et al. 2023). There are many possible solutions to the $a$-distribution mismatch. First, because observational bias favor discovery at smaller $a$, future dedicated large-scale surveys may reveal an intrinsic distribution with more large-$a$ TNOs. Presumably, these surveys could also improve the sample of large-$q$ TNOs, including extreme ones. Alternatively, the 4-Gyr-stable 3:1 resonant TNOs might represent a distinct population formed by resonance sweeping of the disk within ~50 au (see also Footnote 8). For instance, by removing that subpopulation from the observed sample, our model semimajor axis distribution becomes roughly compatible with observations (K-S test probabilities ~5–10%). This idea aligns with the hypothesis that giant planet migration played an important role in populating MMRs beyond 50 au (5:2, 3:1, etc.), as discussed in Section 3.1. Theoretically, the inclusion of grainy giant planet migration (not modeled here) may help populate the region within ~100 au via resonance dropouts from strong n:1 and n:2 resonances in that region. This mechanism may also produce more objects with $q >$ 33 au that would improve the perihelion distribution (Figure 8). Alternatively, perhaps the initial $a$-distribution of the primordial scattered disk contained a higher initial number of objects at smaller $a$ than considered here. In conclusion, future surveys probing large-$a/q/i$, modeling including giant planet migration, and more systematic investigation of the initial semimajor axis distribution of primordial scattered disks are desired to improve the comparison of model results with observations in the distant Kuiper Belt. Furthermore, currently it is not possible to discuss the detections of specific extreme TNOs without a larger sample of such objects and an improved survey simulator that could better constrain "extreme-like" orbits (e.g., only one extreme TNO in Table 1 was found by OSSOS+ surveys: 2013 $SY_{99}$).

After biasing the results obtained from our 10 optimal KBP models, we found that most of them yielded detached populations compatible with observations from OSSOS, as indicated by the perihelion distributions and intrinsic fractions of detected objects in the distant Kuiper Belt (Figure 8 and Table 4). These results also imply that models that included a KBP did much better than the control model in creating detached populations as constrained by observations. In particular, the control model has a substantial deficit of $q > 37$ au detected objects. Also, the detached fraction in that model (~5%) was smaller than the observed fraction (16/125 = 12.8%) and the fractions obtained from the optimal KBP models (~8–12%). In general, we considered KBP models that yielded detached fractions of detected objects at least 80% of that of the observed value as potentially successful. Consistent with the results of Section 3.2.1, these additional results support the suggestion that a resident KBP is required to explain the detached population.

Regarding the analysis of the inclinations of detected and real populations, an important caveat is that the observed high-$i$ fraction of 1.6% (2/125) suffers from critical small-number statistics, so it is impossible to place strict constraints on the results. Nevertheless, we compared the model inclination distributions of our optimal KBP models with observations for completeness. Because OSSOS+ surveys focused on small ecliptic latitudes, the OSS yielded low high-$i$ detections from our model results. Thus, although these models produced significant intrinsic fractions of high-$i$ populations beyond 50 au, the results were consistent with observations after observationally biasing them as constrained by the OSS (Figures 8 and 9). We also found that the optimal KBP models *always* yielded higher high-$i$ detection fractions than the control model (Table 4). If future observations confirm an observed high-$i$ TNO fraction of 1.6% or larger with better statistics, more inclined or more massive KBPs might be required to produce a larger high-$i$ population that could yield biased fractions comparable to the observed.

Finally, five of our optimal KBP models yielded the best results by simultaneously meeting the success criteria of biased detached and high-$i$ fractions (highlighted in bold in Table 4). In particular, the best KBP models produced high-$i$ fractions ≥ 2.5 times the control model, and two of these models yielded ≥ 0.8% fractions that were marginally compatible with the current (limited) observations. These models also produced cumulative distributions of $a$, $q$, and $i$ that were compatible with observations after visual inspection (e.g., Figure 8). Despite the caveats and uncertainties in this analysis, we conclude that the best KBP models can potentially explain the main constraints in the distant Kuiper Belt (Sections 1, 3.1 and 3.2) and be consistent with observations simultaneously (Section 3.3). These final results imply that our proposed KBP, based on the models explored here, should have $i = 30$ deg, $q = 195$ au, and $m = 1.5\ M_\oplus$ ($a = 250$–$350$ au), $2\ M_\oplus$ ($a = 350$–$500$ au) or $3\ M_\oplus$ ($a = 500$ au).

### 3.3.2 Other observational constraints

Could the Earth-like KBP proposed in this work be observable or detectable? The apparent visual magnitude of a KBP depends on its heliocentric distance, albedo, and size. The heliocentric distances of our best KBPs would vary by 195–305 au for the closest hypothetical 'small' orbit ($a_P = 250$ au), 195–505 au for the 'medium' orbit ($a_P = 350$ au), or by as much as 195–805 au for the more distant 'large' orbit ($a_P = 500$ au). For completeness, we illustrate in Figure 10 the apparent sky motions and visual magnitudes of our best KBPs, assuming albedos 0.1 and 0.3 ([9]) and a mean density of 2 g $cm^{-3}$ (e.g., Lykawka 2012). The KBP could be as bright as ~18–21 mag if it is close to the perihelion or has a small orbit. On the other hand, the KBP could also appear darker as ~22–25 mag for most of the time in a more eccentric orbit. Also, a survey must be sensitive to apparent sky motions ~0.46–0.71 arcsec/h and ~0.18–0.71 arcsec/h for the small and large KBP's orbits, respectively. However, our results cannot predict a preferred position in the sky for the location of our best KBPs.

Even without confirmation by observations, under certain conditions the gravitational perturbations of a KBP may be detectable by tracking signals in data obtained by spacecraft. Fienga (2020) and Gomes et al. (2023) focused on the detectability of super-Earths (with 5 or 10 $M_\oplus$) inspired by the Planet 9 model. Although the proposed KBP in our scenario has different orbits and is much less massive, Gomes et al. (2023) also discussed the detectability of less massive hypothetical planets according to the scaling $m_P / d_P{}^3$, where $m_P$ and $d_P$ refer to the planet's mass and heliocentric distance, respectively. Based on the results of that work, KBPs with 1.5 $M_\oplus$, 2 $M_\oplus$, and 3 $M_\oplus$ masses should be detectable in the sky with a 90% probability for a heliocentric distance of less than 298 au, 328 au, and 375 au, respectively. We then determined the probability of detecting our optimal KBPs by considering the fractions of their orbits spent within the critical heliocentric distances above. As expected, the chances of detection are lower for the more distant hypothetical medium and large orbits (Table 4).

## 4 Conclusions

We found that a Kuiper Belt planet (KBP) 1.5–3 times as massive as Earth, located at $a$ = 250–500 au (following an apparent correlation of distance with mass), with an inclined orbit ($i$ = 30 deg) and $q$ = 195 au, can explain the following properties in the distant Kuiper Belt: there must be a prominent population of detached TNOs, well-decoupled from Neptune; a significant fraction of high-$i$ TNOs with $i$ > 45 deg; and a subpopulation of detached or high-$i$ extreme TNOs evolving on peculiar orbits. Furthermore, the proposed KBP is compatible with the existence of long-term Gyr-stable resonant TNOs, and its perturbations do not preclude the formation of scattering populations with $q$ < 40 au. Finally, our scenario's resulting orbital structure beyond ~50 au is reasonably compatible with the inclination and perihelion distributions of OSSOS and other observational constraints. Although our results apparently mismatched OSSOS's semimajor distribution, several possible solutions exist (e.g., considering an improved observational sample, a model including giant planet migration, testing other initial conditions, and so on). Therefore, the KBP model may explain the distant Kuiper Belt in agreement with observations.

In conclusion, the results of the KBP scenario support the existence of a yet-undiscovered planet in the far outer solar system. Furthermore, this scenario also predicts the existence of new TNO populations located beyond 150 au generated by the KBP's perturbations that can serve as observationally testable signatures of the existence of this planet. More detailed knowledge of the orbital structure in the distant Kuiper Belt can reveal or rule out the existence of any hypothetical planet in the outer solar system. The existence of a KBP may also offer new constraints on planet formation and dynamical evolution in the trans-Jovian region.

[9] Sedna is the best representative of a distant and relatively large TNO. Its albedo is 0.32 ± 0.06 (Pal et al. 2012). Also, measurements have revealed that the albedos of detached TNOs lie within ~0.1–0.3 (Santos-Sanz et al. 2012; Farkas-Takacs et al. 2020).

**Acknowledgments**

We thank the anonymous reviewer for the critical and helpful comments, which allowed us to improve the overall presentation of this work. We also thank S. Lawler, L. Peltier, B. Gladman, J.-M. Petit, and M. Bannister for fruitful discussions about the CFEPS L7 model and the OSSOS Survey Simulator. We are also grateful to N. Kaib for discussions and for sharing simulation data. The simulations presented here were partially performed using the general-purpose PC cluster at the Center for Computational Astrophysics in the National Astronomical Observatory of Japan. We are grateful for the generous time allocated to run the simulations. This work was supported by JSPS KAKENHI Grant Number JP20K04049.

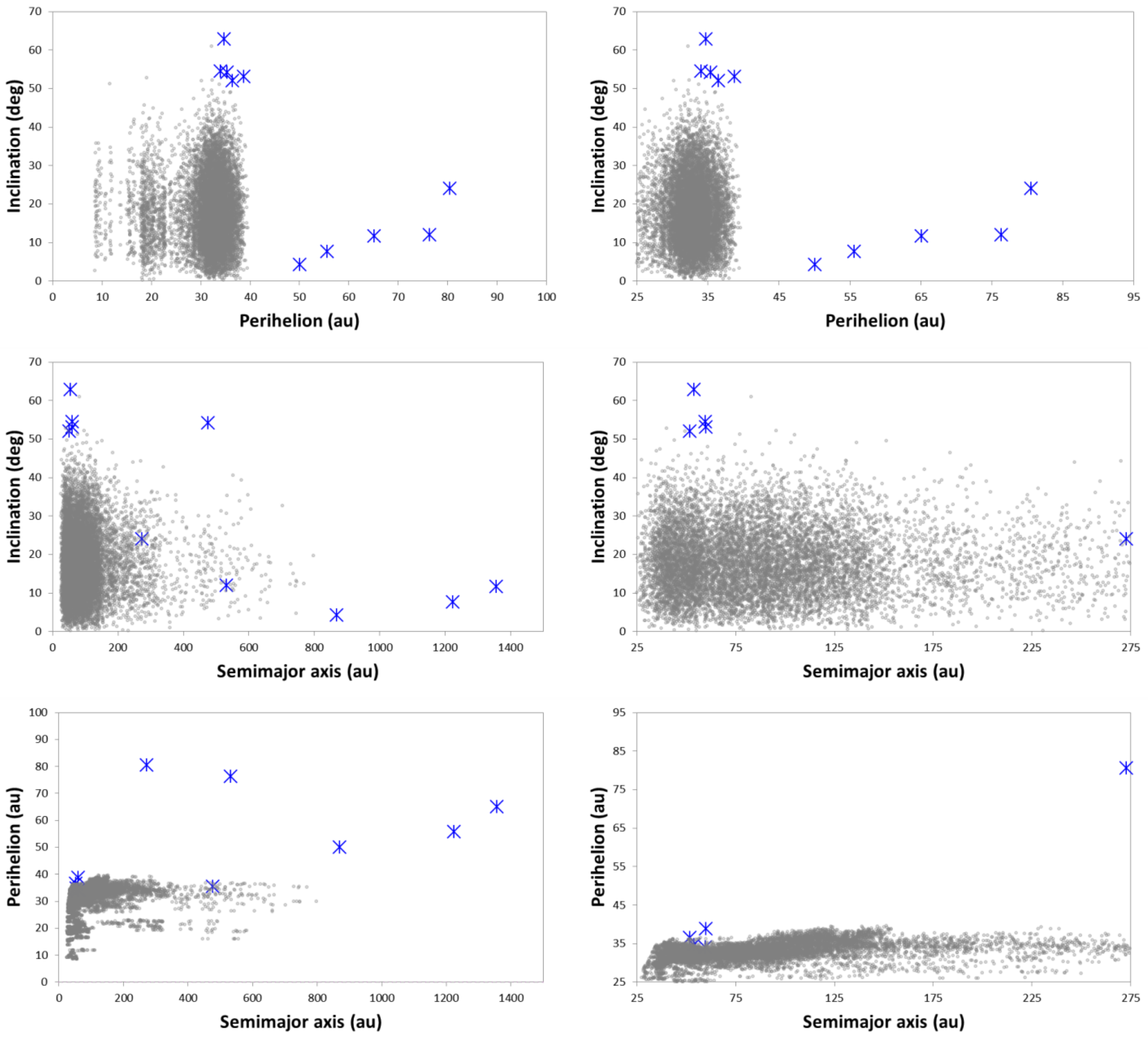

**Figure 1.** Initial conditions of our primordial scattered disk that comprises 48,060 massless particles. Blue asterisks indicate the extreme trans-Neptunian objects (TNOs) discussed in Section 1 and summarized in Table 1.

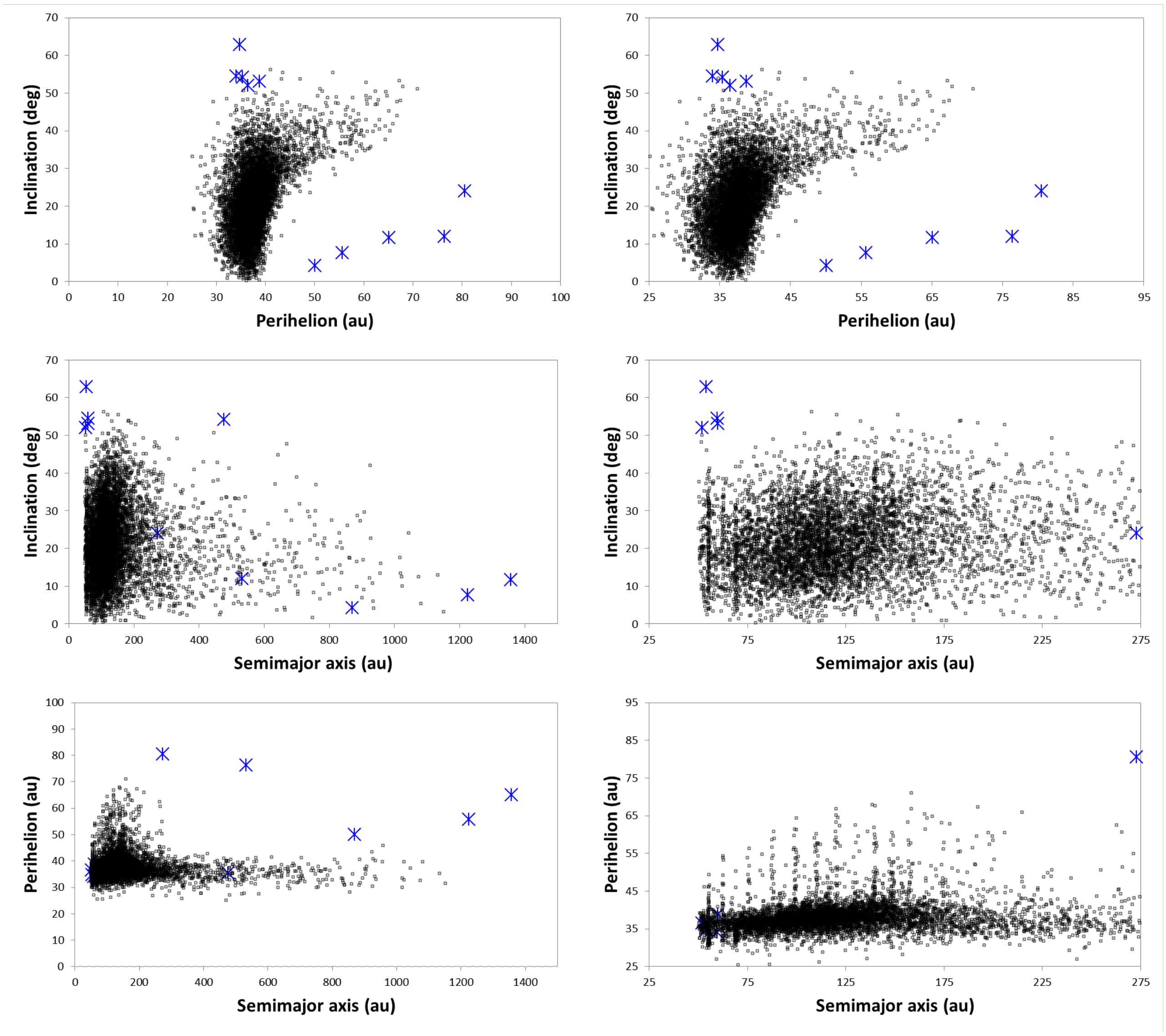

**Figure 2.** Final orbital distribution of the control model after evolving the primordial scattered disk over 4.5 Gyr under the gravitational influence of the four giant planets on their current orbits. Several objects within ~250 au acquire $q$ > 40 au via interactions in strong Neptunian mean motion resonances resulting in a clear $q$–$i$ correlation (top panels). The extreme TNOs (blue asterisks) indicate that the control model cannot explain their orbits. Other shortcomings of this model are discussed in the main text.

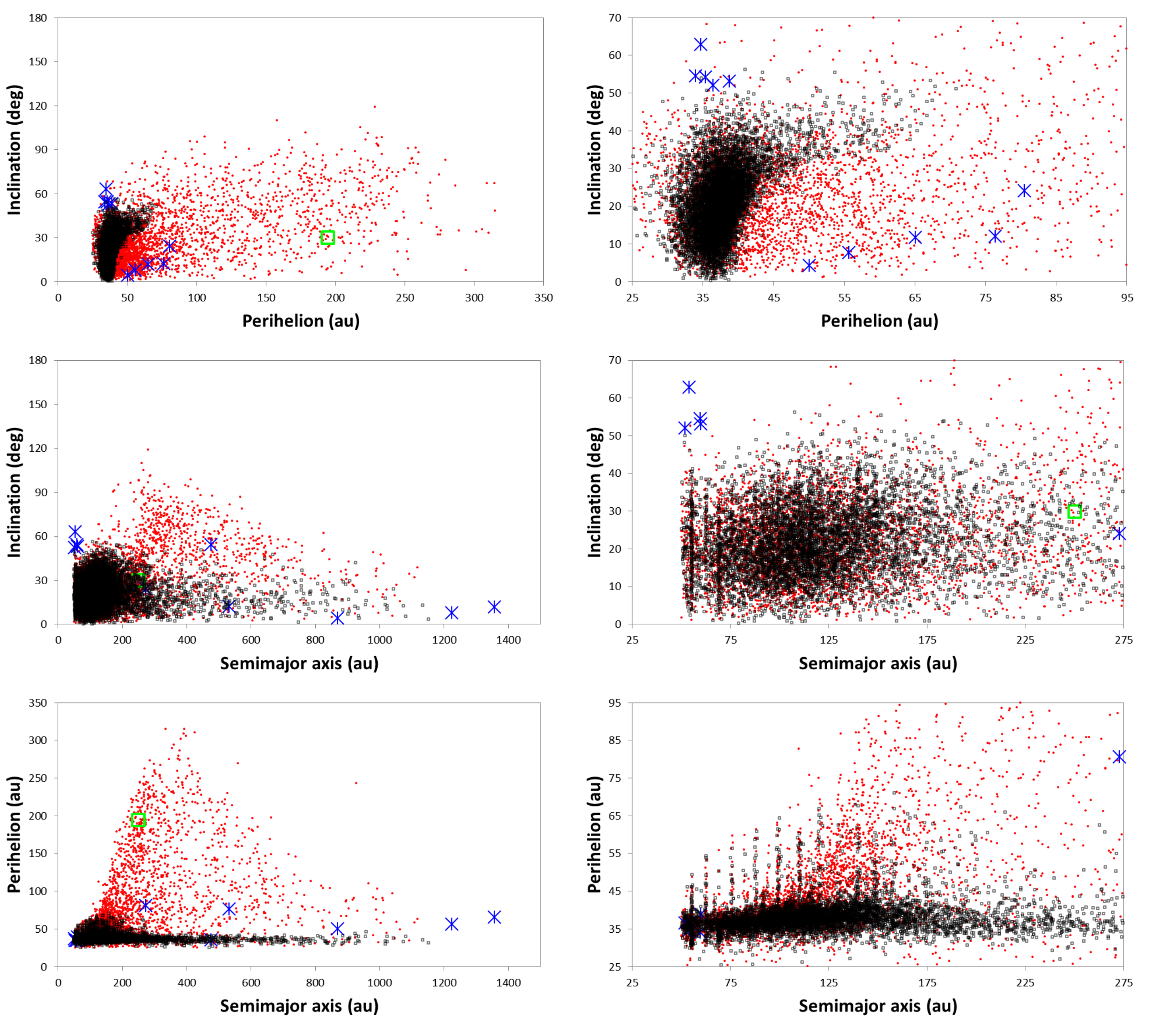

**Figure 3.** Final orbital structure after evolving the primordial scattering population over 4.5 Gyr under the gravitational influence of the four giant planets and a Kuiper Belt planet (KBP) with $m$ = 1.5 $M_{\oplus}$, $a$ = 250 au, $q$ = 195 au, and $i$ = 30 deg. The results obtained for the control model (Figure 2) and this model are represented by black and red symbols, respectively. Blue asterisks represent the extreme TNOs. The KBP's orbit is indicated by the green square.

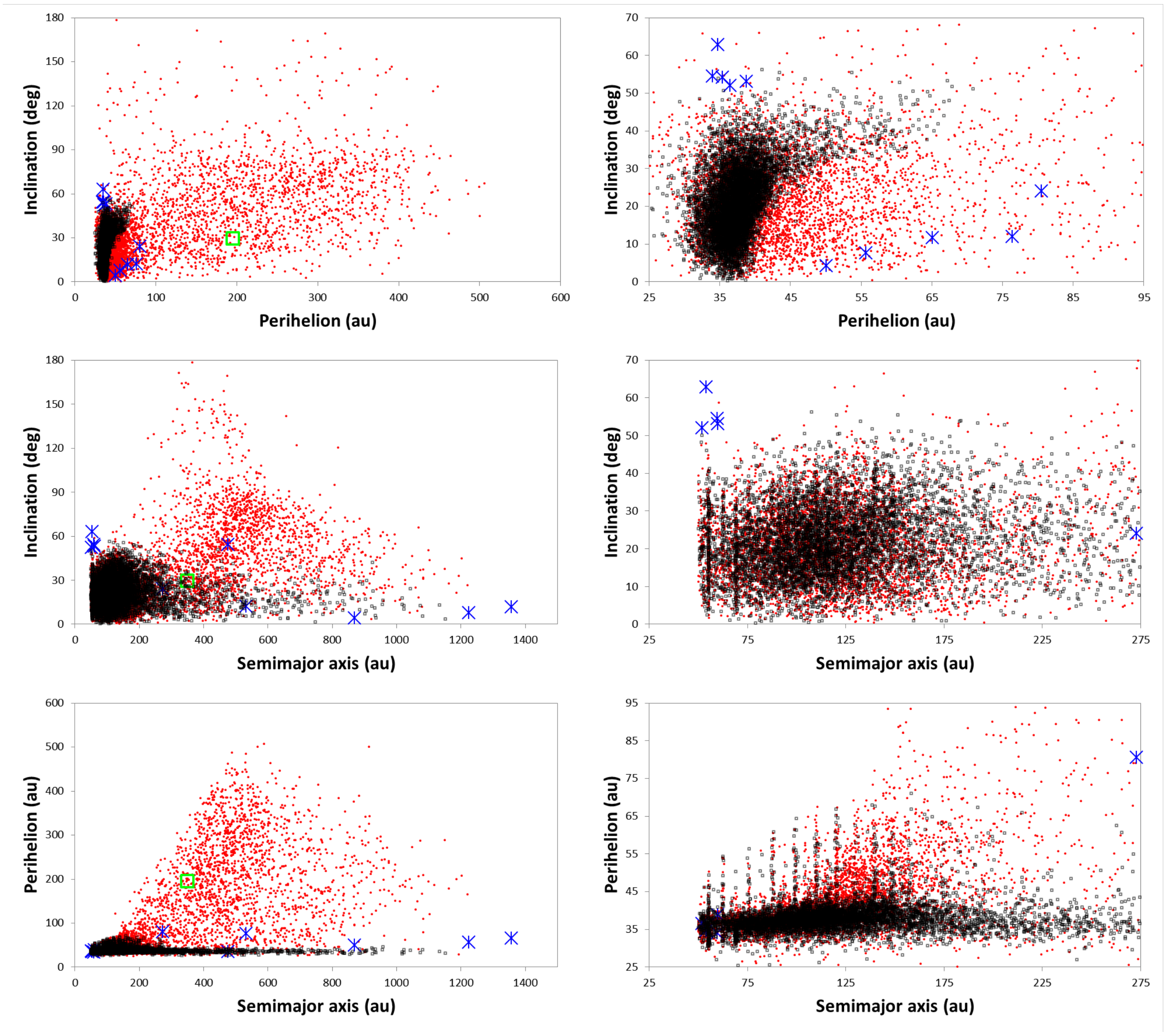

**Figure 4.** Final orbital structure after evolving the primordial scattering population over 4.5 Gyr under the gravitational influence of the four giant planets and a KBP with $m$ = 1.5 $M_{\oplus}$, $a$ = 350 au, $q$ = 195 au, and $i$ = 30 deg. The symbols are explained in the caption of Figure 3.

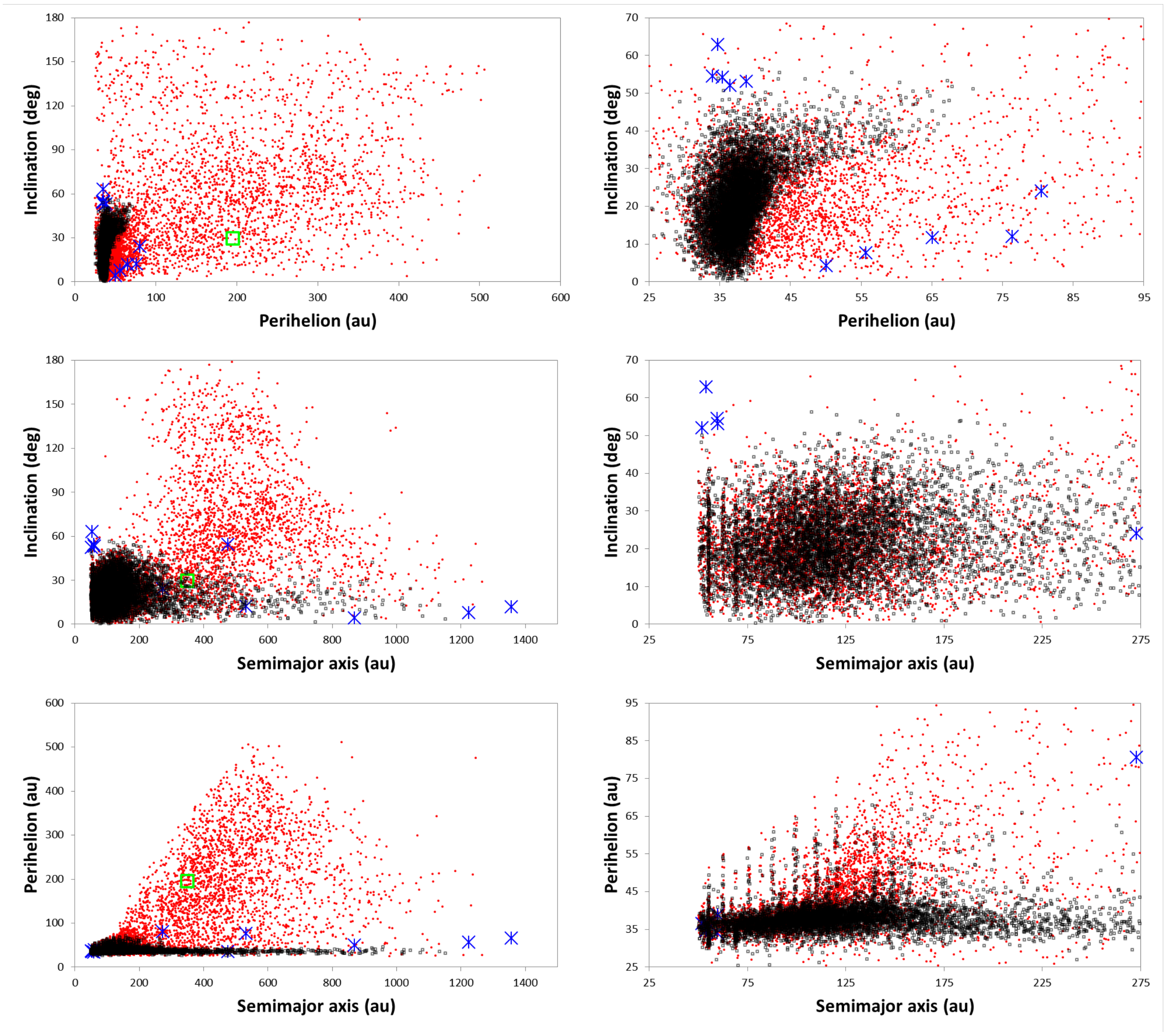

**Figure 5.** Final orbital structure after evolving the primordial scattering population over 4.5 Gyr under the gravitational influence of the four giant planets and a KBP with $m = 2$ M$_\oplus$, $a = 350$ au, $q = 195$ au, and $i = 30$ deg. The symbols are explained in the caption of Figure 3.

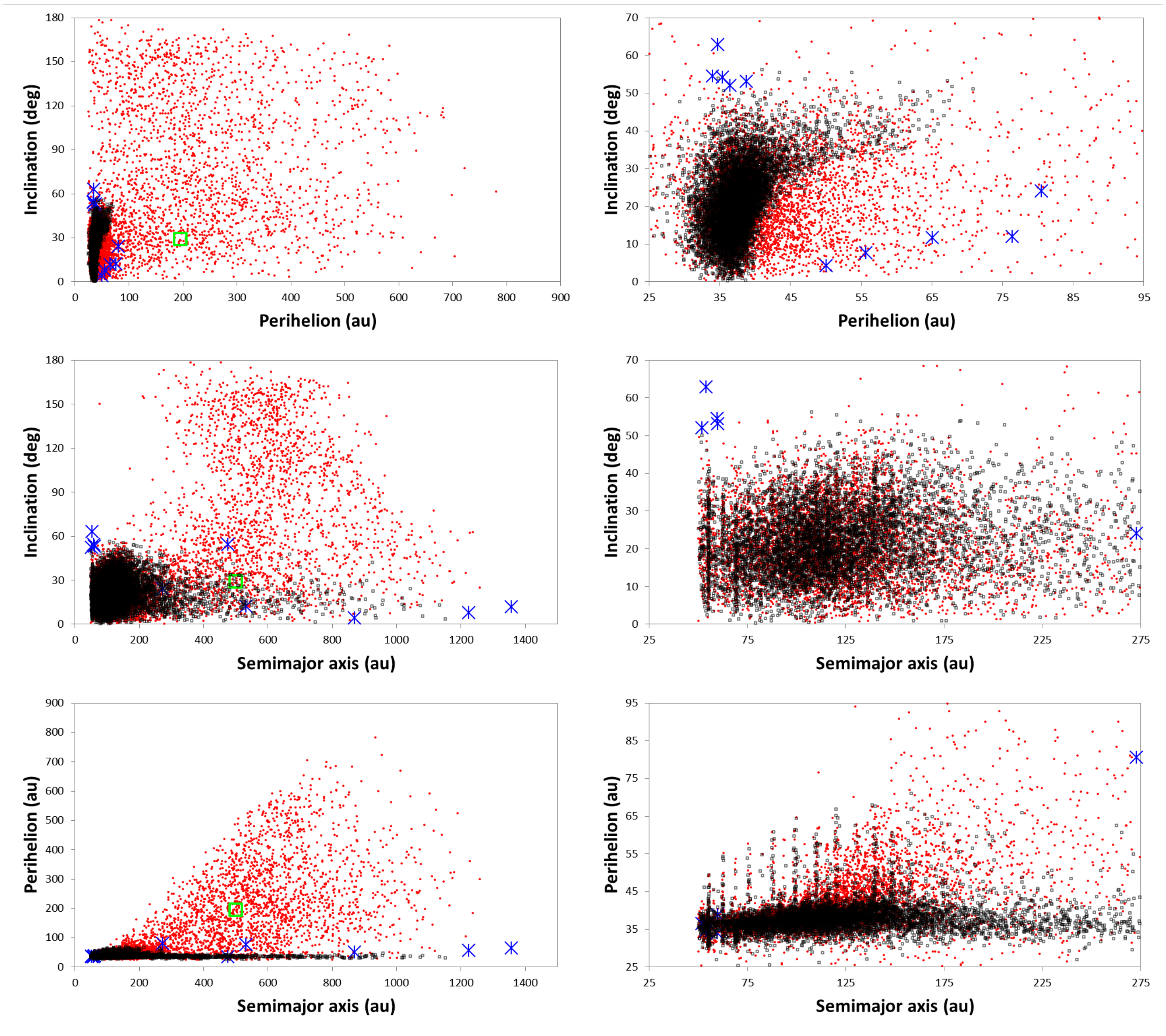

**Figure 6.** Final orbital structure after evolving the primordial scattering population over 4.5 Gyr under the gravitational influence of the four giant planets and a KBP with $m$ = 2 $M_\oplus$, $a$ = 500 au, $q$ = 195 au, and $i$ = 30 deg. The symbols are explained in the caption of Figure 3.

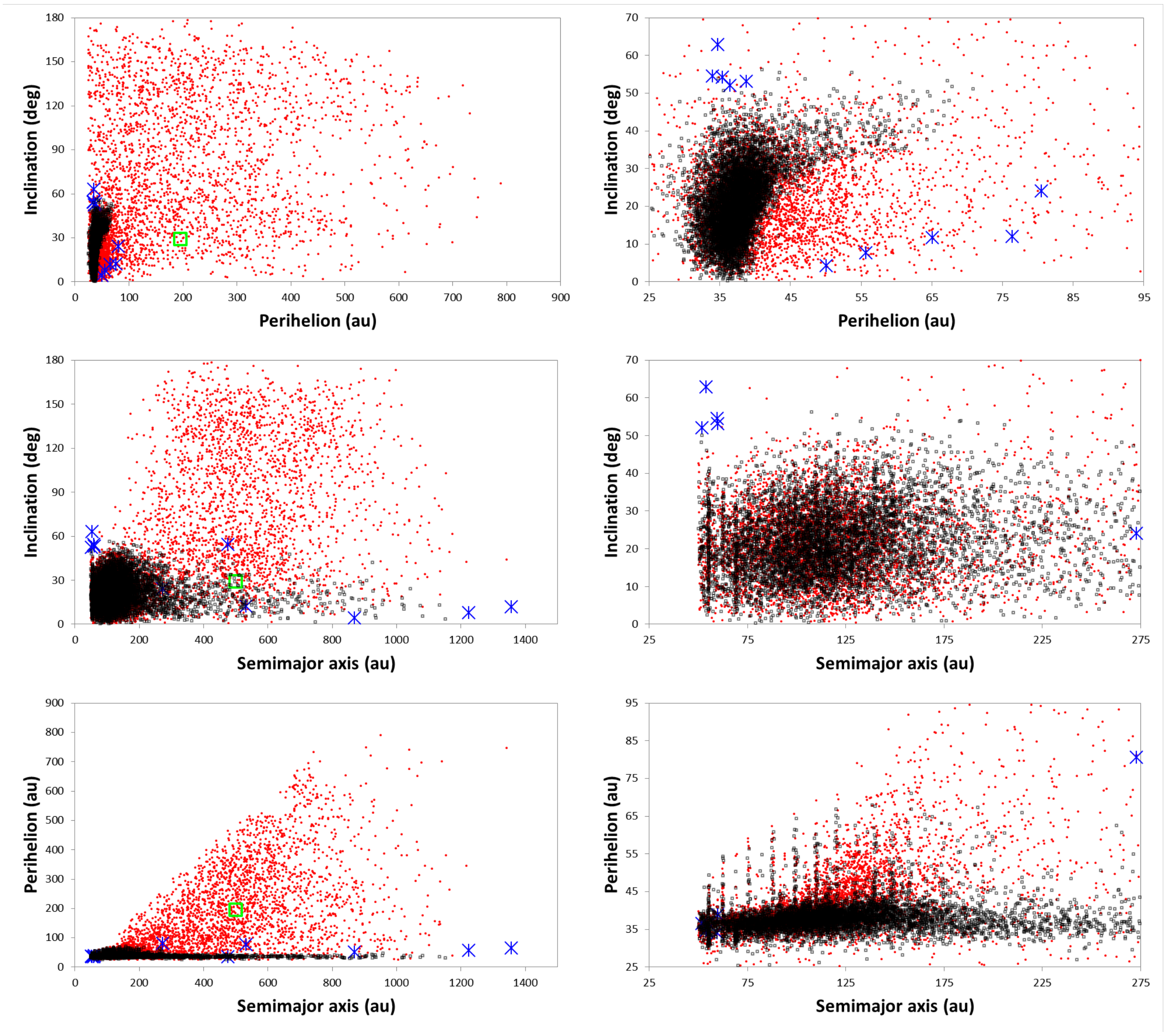

**Figure 7.** Final orbital structure after evolving the primordial scattering population over 4.5 Gyr under the gravitational influence of the four giant planets and a KBP with $m$ = 3 M$_\oplus$, $a$ = 500 au, $q$ = 195 au, and $i$ = 30 deg. The symbols are explained in the caption of Figure 3.

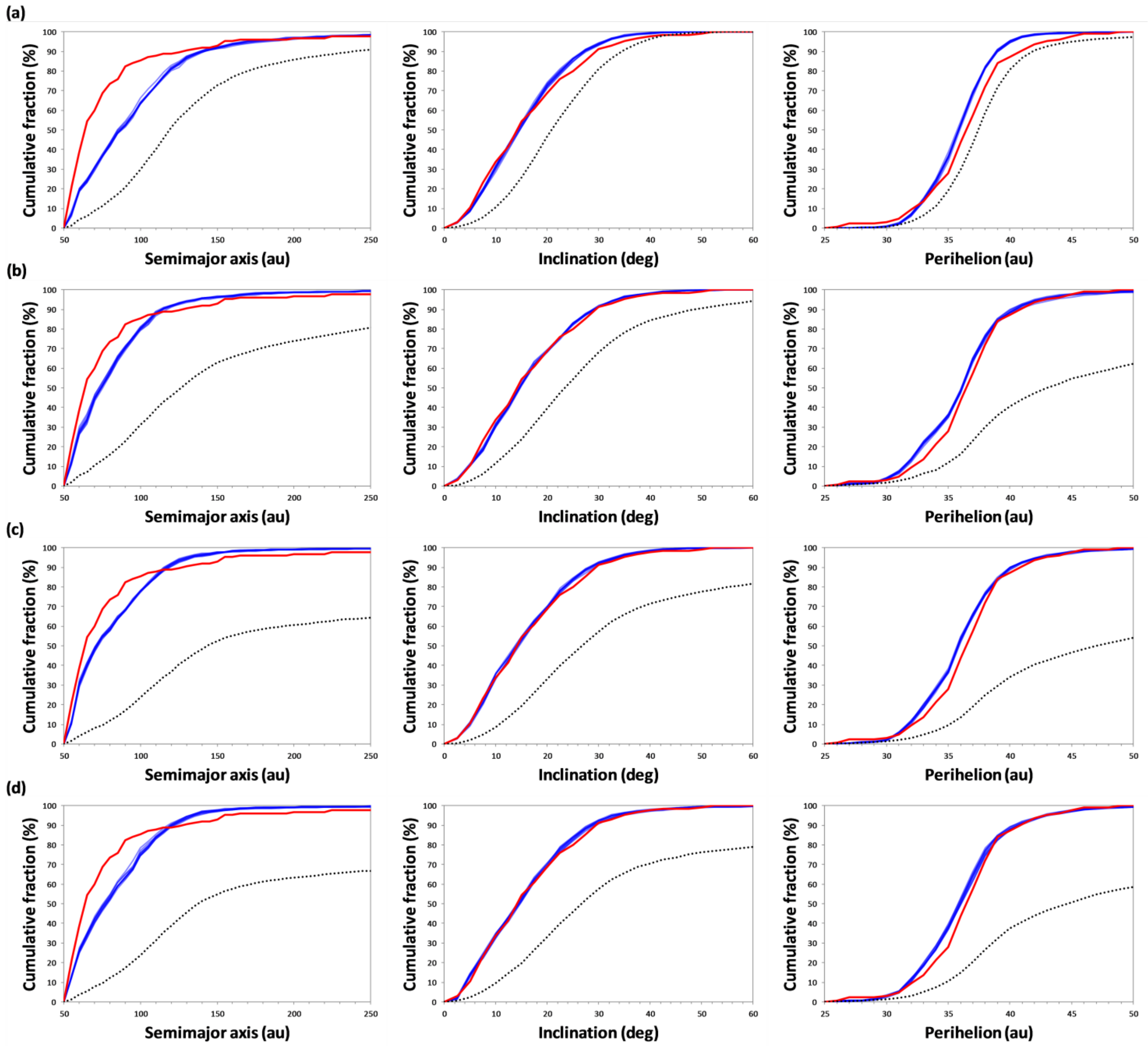

**Figure 8.** Comparison of cumulative orbital distributions of detected objects from representative KBP models after biasing the results (blue curves) and Outer Solar System Origins Survey (OSSOS) observations (red curves). The top three panels (a) show the results for the control model. The other panel-triplets (b–d) show the results for the best KBP models #32, #49, and #59, respectively (Tables 3 and 4). These KBPs had the following properties: $m$ = 1.5 M⊕, $a$ = 250 au (b); $m$ = 2 M⊕, $a$ = 350 au (c); $m$ = 3 M⊕, $a$ = 500 au (d), and the same $q$ = 195 au and $i$ = 30 deg. The models' intrinsic unbiased distributions are indicated by the dotted curves. Only detected objects with $a$ > 50 au and $q$ > 25 au were considered when evaluating the results reported in this figure.

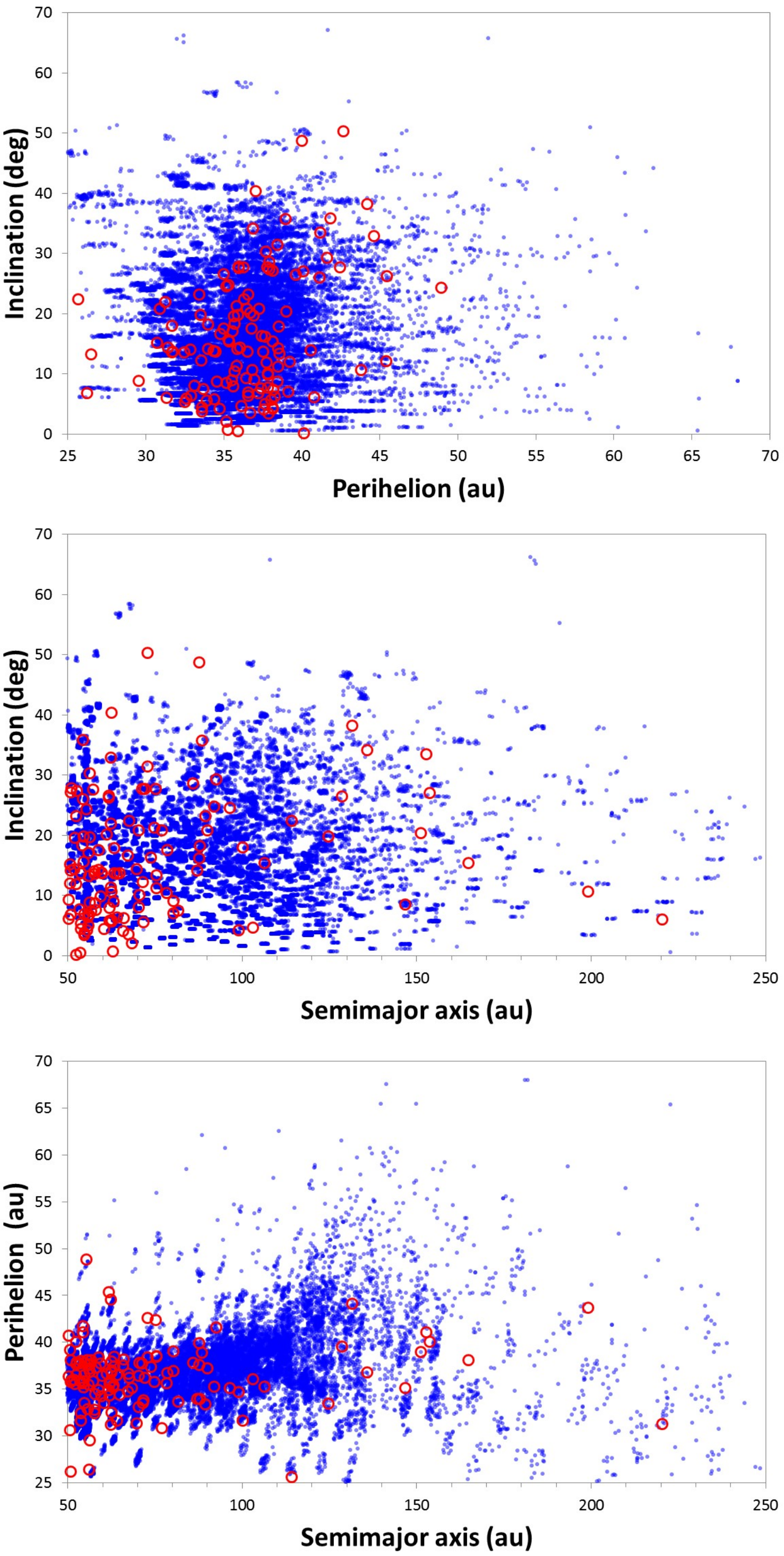

**Figure 9.** Comparison of detected objects from representative KBP model #49 (Tables 3 and 4) with $m = 2$ M$_\oplus$, $a = 350$ au, $q = 195$ au, and $i = 30$ deg (blue symbols) and OSSOS observations (red circles). Only detected objects with $a > 50$ au and $q > 25$ au were considered when evaluating the results reported in this figure.

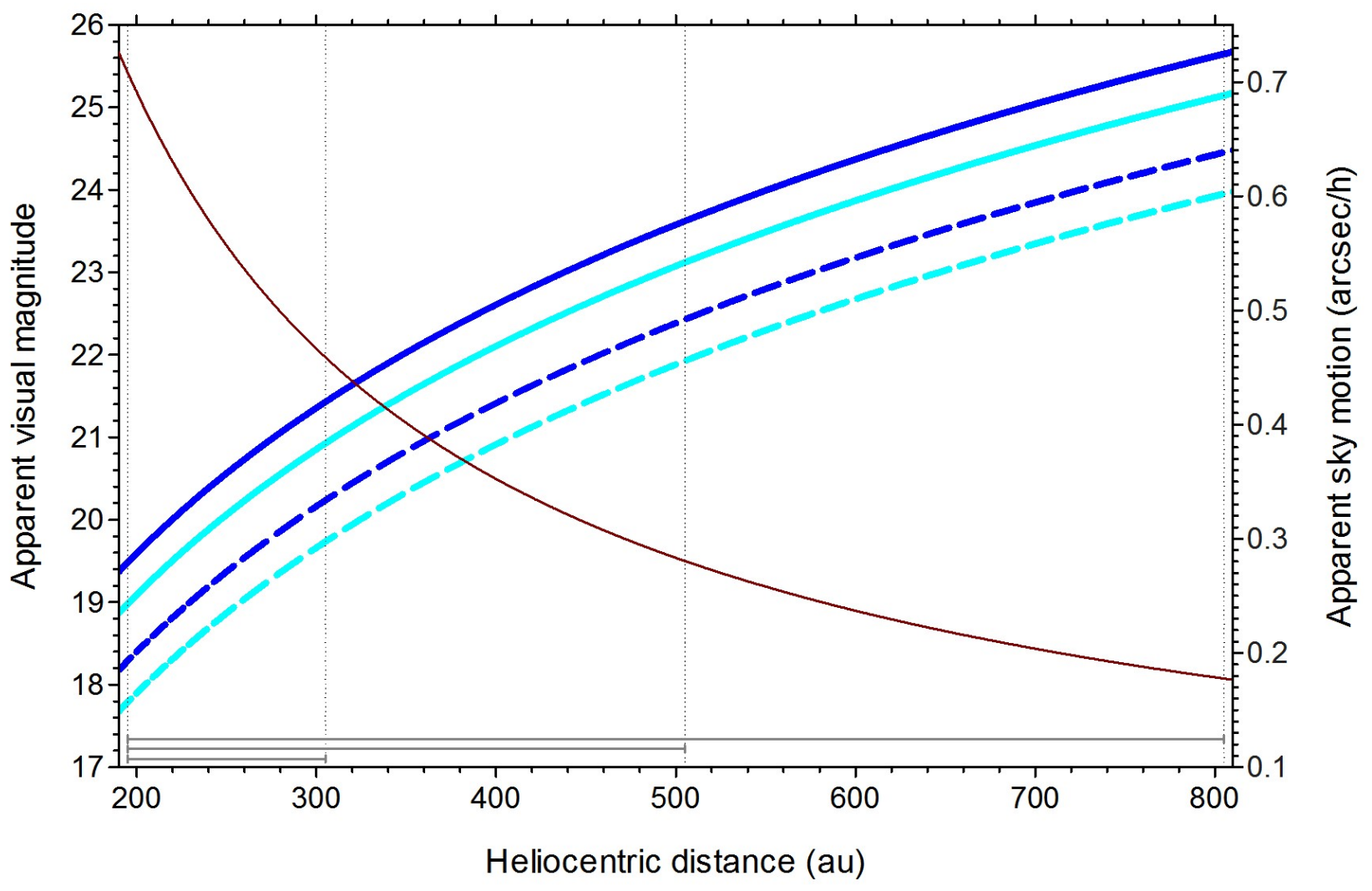

**Figure 10.** Observational constraints on the existence of a KBP based on our best models. The apparent visual magnitudes were estimated for a KBP with $m = 1.5$ M$_\oplus$ (blue curves) and $m = 3$ M$_\oplus$ (cyan curves). The results are shown for an assumed albedo of 0.1 (straight curves) and 0.3 (dashed curves). In all cases, a mean density of 2 g·cm$^{-3}$ was assumed for the KBP (assuming 3 g· cm$^{-3}$ leads to ~0.3 deg darker magnitudes). The variation in heliocentric distance is also indicated by gray bars at the bottom for three hypothetical orbits with $q_P = 195$ au and $a_P = 250$, 350, and 500 au. Finally, the apparent sky motion is indicated by the brown curve. See Section 3 for details.

**Table 1.** Extreme trans-Neptunian objects (TNOs).

| Object | $a$ (au) | $i$ (deg) | $q$ (au) | Parent population | Criterion | Comments | Object reference |
|---|---|---|---|---|---|---|---|
| 2012 $VP_{113}$ | 272.90 | 24.0 | 80.51 | detached | $q > 60$ au | very large perihelion | a |
| (90377) Sedna | 532.43 | 11.9 | 76.38 | detached | $q > 60$ au | idem | b |
| (541132) Leleakuhonua | 1357.43 | 11.7 | 65.13 | detached | $q > 60$ au | idem | c |
| 2013 $SY_{99}$ | 869.59 | 4.2 | 50.06 | detached | $q > 50$ au, $i < 20$ deg | difficult to form via resonant interactions | d |
| 2021 $RR_{205}$ | 1224.28 | 7.6 | 55.66 | detached | $q > 50$ au, $i < 20$ deg | idem | e |
| | | | | | | | |
| 2014 $YX_{91}$ | 54.07 | 62.9 | 34.73 | high-$i$ | $i > 60$ deg | very high inclination | f |
| 2015 $UO_{105}$ | 51.88 | 52.1 | 36.47 | high-$i$ | $i > 50$ deg, $q < 40$ au | difficult to form via resonant interactions | |
| 2017 $FO_{161}$ | 59.58 | 54.4 | 34.04 | high-$i$ | $i > 50$ deg, $q < 40$ au | idem | e |
| 2014 $UN_{225}$ | 59.89 | 53.2 | 38.78 | high-$i$ | $i > 50$ deg, $q < 40$ au | idem | |
| 2015 $BP_{519}$ | 476.20 | 54.1 | 35.38 | high-$i$ | $i > 50$ deg, $q < 40$ au | idem | g |

**Notes.** $a$, $i$, and $q$ represent the TNO's semimajor axis, inclination, and perihelion, respectively. Orbital data retrieved from the AstDys observational database at MJD = 60000 in May 2023. The semimajor axis uncertainties ($da/a$) of Leleakuhonua and 2021 $RR_{205}$ are 9.8% and 2.4%, respectively. See Section 1 for more details. References: a. Trujillo & Sheppard (2014), b. Brown et al. (2004), c. Sheppard et al. (2019), d. Bannister et al. (2017), e. Scott Sheppard Small Body Discoveries (https://sites.google.com/carnegiescience.edu/sheppard/home/discoveries), f. Bernardinelli et al. (2022), g. Becker et al. (2018).

**Table 2.** Summary of resonant TNOs identified that are dynamically stable over Gyr-timescales.

| Mean motion resonance (MMR) | $a_{res}$ (au) | Resonant TNOs after 1 Gyr (4 Gyr) | Aphelion (au) of highest-$e$ resonant TNOs after 1 Gyr (4 Gyr) |
|---|---|---|---|
| 2:1 | 47.8 | 76 (53) | 66 (66) |
| 5:2 | 55.4 | 60 (52) | 80 (81) |
| 3:1 | 62.6 | 15 (12) | 90 (91) |
| 4:1 | 75.8 | 9 (6) | 118 (115) |
| 5:1 | 88.0 | 2 (0) | 141 (—) |
| 6:1 | 99.4 | 1 (1) | 154 (154) |

**Notes.** $a_{res}$ represents the MMR's nominal location. The identified stable resonant populations served as important constraints in simulations investigating the effects of the Kuiper Belt planet with various hypothetical orbits and masses. See the main text for more details.

**Table 3.** A Kuiper Belt planet (KBP) perturbation and its effects on the primordial scattered disk.

| KBP model | $m_P$ ($M_\oplus$) | $a_P$ (au) | $q_P$ (au) | $i_P$ ( deg) | $r_{sca/det}$ | Fhi (%) | $Fhi_{sca}$ (%) | $Fhi_{det}$ (%) | Fi90 (%) | $Fi90_{sca}$ (%) | $Fi90_{det}$ (%) |
|---|---|---|---|---|---|---|---|---|---|---|---|
| *control* | — | — | — | — | 4.18 | 1.2 | 0.4 | 4.5 | 0.0 | 0.0 | 0.0 |
| *1* | 0.1 | 250 | 95 | 1 | 1.98 | 1.1 | 0.3 | 2.6 | 0.0 | 0.0 | 0.0 |
| *2* | 0.1 | 250 | 95 | 30 | 2.20 | 1.3 | 0.4 | 3.4 | 0.0 | 0.0 | 0.0 |
| *3* | 0.1 | 350 | 65 | 1 | 1.89 | 1.3 | 0.4 | 3.1 | 0.0 | 0.0 | 0.0 |
| *4* | 0.1 | 350 | 65 | 30 | 2.42 | 1.3 | 0.4 | 3.4 | 0.0 | 0.0 | 0.0 |
| *5* | 0.1 | 350 | 95 | 1 | 2.56 | 1.1 | 0.3 | 3.4 | 0.0 | 0.0 | 0.0 |
| *6* | 0.1 | 350 | 95 | 30 | 3.02 | 1.2 | 0.5 | 3.3 | 0.0 | 0.0 | 0.0 |
| *7* | 0.1 | 500 | 65 | 1 | 2.47 | 1.2 | 0.3 | 3.3 | 0.0 | 0.0 | 0.0 |
| *8* | 0.1 | 500 | 65 | 30 | 3.11 | 1.6 | 0.5 | 5.0 | 0.0 | 0.0 | 0.0 |
| *9* | 0.1 | 500 | 95 | 1 | 3.07 | 1.3 | 0.4 | 4.1 | 0.0 | 0.0 | 0.0 |
| *10* | 0.1 | 500 | 95 | 30 | 3.38 | 1.2 | 0.5 | 3.7 | 0.0 | 0.0 | 0.0 |
| *11* | 0.3 | 250 | 120 | 1 | 1.03 | 0.9 | 0.6 | 1.3 | 0.0 | 0.0 | 0.0 |
| *12* | 0.3 | 250 | 120 | 30 | 1.18 | 1.7 | 0.6 | 3.0 | 0.0 | 0.0 | 0.0 |
| *13* | 0.3 | 350 | 120 | 1 | 1.28 | 1.1 | 0.4 | 2.0 | 0.0 | 0.0 | 0.0 |
| *14* | 0.3 | 350 | 120 | 30 | 1.57 | 2.0 | 0.6 | 4.0 | 0.0 | 0.0 | 0.0 |
| *15* | 0.3 | 500 | 120 | 1 | 1.80 | 0.9 | 0.3 | 2.0 | 0.0 | 0.0 | 0.0 |
| *16* | 0.3 | 500 | 120 | 30 | 2.09 | 2.0 | 0.5 | 4.9 | 0.0 | 0.0 | 0.0 |
| *17* | 0.5 | 250 | 145 | 1 | 0.87 | 0.8 | 0.5 | 1.0 | 0.0 | 0.0 | 0.0 |
| *18* | 0.5 | 250 | 145 | 30 | 0.90 | 2.6 | 0.6 | 4.4 | 0.0 | 0.0 | 0.0 |
| *19* | 0.5 | 350 | 145 | 1 | 0.97 | 0.9 | 0.6 | 1.1 | 0.0 | 0.0 | 0.0 |
| *20* | 0.5 | 350 | 145 | 30 | 1.18 | 3.0 | 0.7 | 5.7 | <0.1 | 0.0 | <0.1 |
| *21* | 0.5 | 500 | 145 | 1 | 1.22 | 0.8 | 0.4 | 1.2 | 0.0 | 0.0 | 0.0 |
| *22* | 0.5 | 500 | 145 | 30 | 1.53 | 3.3 | 0.5 | 7.5 | 0.0 | 0.0 | 0.0 |
| *23* | 1.0 | 250 | 195 | 1 | 1.00 | 1.3 | 0.5 | 2.0 | 0.0 | 0.0 | 0.0 |
| *24* | 1.0 | 250 | 195 | 30 | 0.85 | 3.3 | 0.5 | 5.7 | <0.1 | 0.0 | <0.1 |
| *25* | 1.0 | 350 | 195 | 1 | 0.82 | 1.1 | 0.8 | 1.4 | <0.1 | 0.0 | <0.1 |
| *26* | 1.0 | 350 | 195 | 30 | 0.92 | 7.7 | 0.7 | 14.1 | 0.2 | 0.0 | 0.3 |
| *27* | 1.0 | 350 | 245 | 1 | 1.46 | 1.1 | 0.4 | 2.0 | 0.0 | 0.0 | 0.0 |
| *28* | 1.0 | 350 | 245 | 30 | 1.30 | 1.9 | 0.5 | 3.8 | <0.1 | 0.0 | <0.1 |
| *29* | 1.0 | 500 | 195 | 1 | 1.10 | 1.4 | 0.7 | 2.2 | <0.1 | 0.0 | <0.1 |
| *30* | 1.0 | 500 | 195 | 30 | 1.00 | 9.6 | 0.4 | 18.8 | 1.1 | 0.0 | 2.2 |
| *31* | 1.5 | 250 | 195 | 15 | 0.62 | 7.1 | 0.7 | 11.1 | <0.1 | 0.0 | <0.1 |
| *32* | **1.5** | **250** | **195** | **30** | **0.68** | **12.1** | **1.3** | **19.5** | **0.5** | **0.0** | **0.8** |
| *33* | 1.5 | 350 | 195 | 1 | 0.76 | 2.6 | 0.7 | 4.0 | 0.1 | 0.0 | 0.2 |
| *34* | **1.5** | **350** | **195** | **15** | **0.67** | **10.2** | **1.4** | **16.1** | **0.2** | **<0.1** | **0.3** |
| *35* | **1.5** | **350** | **195** | **30** | **0.64** | **17.4** | **1.4** | **27.6** | **2.1** | **0.1** | **3.4** |
| *36* | 1.5 | 500 | 195 | 1 | 0.95 | 3.7 | 0.7 | 6.6 | 1.0 | <0.1 | 2.0 |
| *37* | **1.5** | **500** | **195** | **15** | **0.82** | **10.0** | **1.2** | **17.2** | **3.7** | **0.5** | **6.3** |
| *38* | **1.5** | **500** | **195** | **30** | **0.74** | **16.5** | **1.2** | **27.9** | **6.4** | **0.3** | **11.0** |
| *39* | 1.5 | 500 | 245 | 1 | 1.13 | 4.3 | 0.8 | 8.1 | 0.6 | <0.1 | 1.3 |
| *40* | 1.5 | 500 | 245 | 15 | 1.11 | 6.1 | 0.4 | 12.3 | 1.1 | <0.1 | 2.2 |
| *41* | 1.5 | 500 | 245 | 30 | 0.99 | 11.2 | 0.9 | 21.4 | 1.4 | <0.1 | 2.7 |
| *42* | 2.0 | 250 | 195 | 1 | 0.60 | 2.9 | 0.9 | 4.1 | <0.1 | 0.0 | <0.1 |
| *43* | **2.0** | **250** | **195** | **15** | **0.49** | **13.0** | **1.2** | **18.8** | **0.2** | **0.0** | **0.3** |
| *44* | **2.0** | **250** | **195** | **30** | **0.53** | **22.0** | **1.8** | **32.7** | **2.5** | **0.1** | **3.7** |
| *45* | 2.0 | 250 | 245 | 1 | 1.44 | 2.2 | 0.5 | 4.5 | 0.0 | 0.0 | 0.0 |
| *46* | 2.0 | 250 | 245 | 30 | 0.94 | 3.6 | 0.9 | 6.2 | <0.1 | 0.0 | <0.1 |
| *47* | **2.0** | **350** | **195** | **1** | **0.73** | **5.5** | **1.3** | **8.6** | **0.5** | **<0.1** | **0.9** |
| *48* | **2.0** | **350** | **195** | **15** | **0.64** | **13.6** | **2.1** | **21.0** | **1.6** | **0.3** | **2.4** |
| *49* | **2.0** | **350** | **195** | **30** | **0.52** | **25.0** | **2.8** | **36.5** | **8.4** | **1.1** | **12.2** |
| *50* | 2.0 | 350 | 245 | 1 | 0.81 | 3.5 | 0.5 | 6.0 | <0.1 | 0.0 | <0.1 |
| *51* | **2.0** | **350** | **245** | **30** | **0.72** | **13.6** | **1.1** | **22.5** | **1.0** | **0.0** | **1.7** |
| *52* | **2.0** | **500** | **195** | **1** | **0.86** | **8.4** | **2.2** | **13.6** | **3.6** | **0.6** | **6.2** |
| *53* | **2.0** | **500** | **195** | **15** | **0.74** | **13.9** | **2.0** | **22.7** | **7.2** | **0.7** | **12.1** |
| *54* | **2.0** | **500** | **195** | **30** | **0.65** | **21.2** | **2.4** | **33.4** | **11.3** | **0.9** | **18.1** |
| *55* | 2.0 | 500 | 245 | 1 | 1.04 | 9.2 | 1.1 | 17.6 | 3.9 | 0.4 | 7.6 |
| *56* | **2.0** | **500** | **245** | **15** | **0.95** | **12.4** | **1.4** | **22.9** | **4.8** | **0.3** | **9.1** |
| *57* | **2.0** | **500** | **245** | **30** | **0.86** | **18.2** | **1.4** | **32.5** | **7.6** | **0.7** | **13.6** |
| *58* | **3.0** | **500** | **195** | **1** | **0.71** | **14.6** | **2.5** | **23.2** | **7.6** | **1.2** | **12.1** |
| *59* | **3.0** | **500** | **195** | **30** | **0.58** | **26.3** | **4.0** | **39.3** | **14.7** | **2.0** | **22.2** |
| *60* | **3.0** | **500** | **245** | **1** | **0.87** | **16.1** | **1.9** | **28.5** | **9.9** | **1.0** | **17.6** |
| *61* | **3.0** | **500** | **245** | **30** | **0.72** | **25.5** | **2.8** | **41.9** | **14.1** | **1.4** | **23.3** |
| Intrinsic fractions | — | — | — | — | ≤1.0 | ≥ 2.0 | ≥ 1.0 | ≥ 6.0 | > 0 | — | — |

**Notes.** $r_{sca/det}$ is the ratio of scattering and detached populations. Fhi and Fi90 represent the fraction of high-*i* ($i > 45$ deg) and retrograde ($i > 90$ deg) objects. The subscripts 'sca' and 'det' refer to results for the scattering and detached populations, respectively. The first row shows the control results obtained in the control model, which considered only the four giant planets and their current orbits. The KBP model considered the giant planets and an additional resident planet with various orbits and masses. Objects with $a > 50$ au and $q > 25$ au were considered when evaluating the results reported in this table. Favored KBP models simultaneously satisfied the estimated intrinsic fractions (highlighted in bold).

**Table 4.** Properties of an observationally biased distant trans-Neptunian region based on our optimal KBP models.

| KBP model | $m_P$ ($M_\oplus$) | $a_P$ (au) | $q_P$ (au) | $i_P$ (deg) | <$Fhi_{biased}$> (%) | <$F_{det,biased}$> (%) | Comments | Probability of detection (%) |
|---|---|---|---|---|---|---|---|---|
| *control* | — | — | — | — | 0.2 (0.2) | 4.9 (5.3) | | |
| *32* | **1.5** | **250** | **195** | **30** | **0.7 (0.6)** | **11.4 (12.2)** | best | 72 |
| *35* | **1.5** | **350** | **195** | **30** | **0.5 (0.5)** | **11.9 (12.4)** | best | 23 |
| *38* | 1.5 | 500 | 195 | 30 | 0.3 (0.3) | 10.3 (10.8) | | 17 |
| *44* | 2.0 | 250 | 195 | 30 | 0.6 (0.8) | 9.3 (10.0) | (best?) | 90 |
| *49* | **2.0** | **350** | **195** | **30** | **0.5 (0.6)** | **10.2 (11.3)** | best | 28 |
| *51* | 2.0 | 350 | 245 | 30 | 0.3 (0.3) | 9.4 (10.6) | | 31 |
| *54* | **2.0** | **500** | **195** | **30** | **0.8 (0.8)** | **11.8 (12.1)** | best | 19 |
| *57* | 2.0 | 500 | 245 | 30 | 0.3 (0.4) | 8.2 (8.7) | | 19 |
| *59* | **3.0** | **500** | **195** | **30** | **1.3 (1.3)** | **11.9 (12.2)** | best | 23 |
| *61* | 3.0 | 500 | 245 | 30 | 1.0 (1.0) | 8.2 (8.9) | | 23 |
| | | | | | | | | |
| Success criteria | — | — | — | — | ≥ 0.5 | ≥ 10.2 | | |
| Observations | — | — | — | — | 1.6 | 12.8 | | |

**Notes.** KBP models that satisfied all the distant Kuiper Belt's main constraints and other conditions discussed in the text represent our optimal models. <$Fhi_{biased}$> and <$F_{det,biased}$> are the fractions of high-*i* and detached objects detected by the Outer Solar System Origins Survey (OSSOS) Survey Simulator, respectively. The results were obtained for absolute magnitudes described by the divot and knee (within parentheses) distributions as favored in the literature. Other variables and model definitions are explained in Table 3's caption. Only detected objects with $a$ > 50 au and $q$ > 25 au were considered when evaluating the results reported in this table. This restriction yielded ~3,000 detected objects per model and 125 TNOs characterized by OSSOS+ surveys. Among the observed TNOs, 2 and 16 are high-*i* and detached, yielding 2/125 = 1.6% and 16/125 = 12.8%, respectively. The probability of detection is defined by the fraction of the KBP's orbit spent within the detectability distance as constrained by spacecraft tracking multiplied by a 90% probability of detectability in the sky, as determined by Gomes et al. (2023). Our best KBPs highlighted in bold satisfied the detached and high-*i* detected populations' success criteria. See Section 3.3.1 for more details.

## Appendix

Table A1 lists the unstable TNOs with q = 15–25 au and high-i orbits. Tables A2 and A3 provide details about the perturbation of a KBP on a given stable resonant population located beyond 47 au. See also Sections 1 and 2.

**Table A1.** Unstable trans-Neptunian objects (TNOs) with $q$ = 15–25 au and $i$ > 45 deg.

| Object | $a$ (au) | $i$ (deg) | $q$ (au) |
|---|---|---|---|
| 2015 $AD_{298}$ | 426.10 | 47.2 | 18.77 |
| 2013 $TB_{187}$ | 78.31 | 50.8 | 16.45 |
| 2012 $UW_{177}$ | 30.19 | 53.8 | 22.28 |
| (523700) 2014 $GM_{54}$ | 39.50 | 54.2 | 24.20 |
| 2010 $WG_9$ | 53.57 | 70.2 | 18.73 |
| (127546) 2002 $XU_{93}$ | 65.88 | 77.9 | 20.95 |
| (523719) 2014 $LM_{28}$ | 284.60 | 84.8 | 16.78 |
| (528219) 2008 $KV_{42}$ | 42.05 | 103.4 | 21.16 |
| (471325) 2011 $KT_{19}$ | 35.75 | 110.3 | 23.90 |
| 2021 $TH_{165}$ | 39.40 | 154.9 | 21.55 |
| 2020 $YR_3$ | 443.52 | 169.3 | 16.49 |

**Notes.** $a$, $i$, and $q$ represent the TNO's semimajor axis, inclination, and perihelion, respectively. Orbital data retrieved from the AstDys observational database at MJD = 60000 in May 2023. See Sections 1 and 3.2.2 for discussions. See also Gladman et al. (2009), Brasser et al. (2012) and Chen et al. (2016) for more details on these objects.

**Table A2.** Summary of a Kuiper Belt planet (KBP) perturbation in the Gyr-stable resonant populations located beyond 47 au.

| $m_P$ ($M_\oplus$) | $a_P$ (au) | $q_P$ (au) | Stable resonant populations tested | Affected populations and distributions (observed TNOs) | Affected populations and distributions (L7 objects) | Result |
|---|---|---|---|---|---|---|
| 0.1 | 100 | 65 | real+61, L7 | **e52,** He52, n52, **n61, r5/2** | **e21, He21, e52,** n52, n31, n51, **r5/2, r5/3** | BAD |
| 0.1 | 100 | 80 | real+61, L7 | e52, n31, **n41**, **n61,** r5/3 | e31, e51, n51, r3/5, **r5/3** | BAD |
| 0.1 | 100 | 95 | real+61, L7 | **n61** | e51, n51, **r5/3** | BAD |
| 0.1 | 105 | 105 | 61, L7#51 | **n61** | e51, n51 | BAD |
| 0.1 | 110 | 110 | 61, L7#51 | **n61** | e51, n51 | BAD |
| 0.1 | 115 | 115 | real+61 | **n41, n61** | — | BAD |
| 0.1 | 120 | 120 | 61, L7#51 | **n61** | n51 | possibly bad |
| 0.1 | 140 | 115 | real+61 | **n61** | — | marginally ok |
| 0.1 | 140 | 140 | real+61 | **n61** | — | marginally ok |
| 0.1 | 175 | 65 | real+61 | He21, e52, n31, **n61**, r5/2, r3/5 | — | BAD |
| 0.1 | 175 | 80 | real+61 | **n61?**, **r3/5** | — | possibly bad |
| 0.1 | 175 | 95 | real+61, L7#51 | **n61?** | | ok? |
| 0.1 | 175 | 145 | real+61 | **n61?** | — | ok? |
| 0.1 | 250 | 65 | real+61, L7 | **n61** | He21, **e52** | possibly bad |
| 0.1 | 250 | 80 | real+61, L7 | **n61?** | | ok? |
| 0.1 | 250 | 95 | real+61, L7 | **n61?** | | ok? |
| 0.1 | 250 | 145 | 61, L7 | **n61?** | | ok? |
| 0.1 | 500 | 145 | 61, L7 | | | **OK** |
| | | | | | | |
| 0.3 | 100 | 95 | real+61 | *n31*, **n41**, **n61, r3/5** | — | BAD |
| 0.3 | 110 | 110 | 61, L7#51 | **n61** | **e51, n51** | BAD |
| 0.3 | 115 | 115 | 61, L7 | **n61** | **e51, n51, r5/3** | BAD |
| 0.3 | 120 | 120 | 61, L7#51 | **n61** | **e51, n51** | BAD |
| 0.3 | 140 | 115 | 61, L7 | **n61** | **e51, n51, r5/3** | BAD |
| 0.3 | 140 | 140 | real+61 | **n61** | — | marginally ok |
| 0.3 | 175 | 65 | real+61 | **He21,** e52, He52, n21, **n52,** n31, **n41**, **n61, r5/2** | — | BAD |
| 0.3 | 175 | 80 | real+61 | e52, He52, **n61, r5/2** | — | BAD |
| 0.3 | 175 | 95 | real+61 | n41, **n61** | — | possibly bad |
| 0.3 | 175 | 120 | 61, L7#51 | **n61** | n51 | possibly bad |
| 0.3 | 175 | 145 | real+61 | **n61** | — | marginally ok |
| 0.3 | 250 | 65 | real+61 | He21, e52, He52, n52, **n61, r5/2** | — | BAD |
| 0.3 | 250 | 80 | real+61 | *n41,* **n61, r5/2,** r3/5 | — | BAD |
| 0.3 | 250 | 95 | real+61 | n41, **n61** | — | BAD |
| 0.3 | 250 | 120 | 61, L7#51 | **n61** | | marginally ok |
| 0.3 | 500 | 120 | 61, L7#51 | **n61?** | | ok? |
| | | | | | | |
| 0.5 | 175 | 65 | real, L7 | e21, He21, *e52*, **He52**, n21, **n52, n31, n41, r5/2** | **e21, He21, e52, He52, e31,** e51, n21**, n52, n31, n51, r5/2**, r3/5**, r5/3** | BAD |
| 0.5 | 175 | 80 | real, L7 | e52, He52, **n31**, **n41**, *r5/2*, r3/5 | He21, **e52**, **He52**, **e31**, e51, n31, **n51**, **r3/5, r5/3** | BAD |
| 0.5 | 175 | 95 | real+61, L7 | **n41, n61,** r5/3 | **e51**, **n51**, **r5/3** | BAD |
| 0.5 | 175 | 145 | real+61 | **n61** | — | marginally ok |
| 0.5 | 250 | 65 | real, L7 | **He21,** *e52*, **He52**, n21, **n52, n41, r5/2** | **e21, He21, e52**, He52, e31, e51, n21, **n52, n31, n51**, **r5/2,** | BAD |
| | | | | | **r5/3** | |
| 0.5 | 250 | 80 | real, L7 | e52, **He52,** n52, n31, *n41*, r3/5, **r5/2** | **e52**, He52, **e31**, e51, **n51**, r5/2, **r3/5, r5/3** | BAD |
| 0.5 | 250 | 95 | real+61, L7 | n41, **n61** | **n51, r5/3** | BAD |
| 0.5 | 250 | 145 | 61, L7 | **n61** | | marginally ok |
| 0.5 | 500 | 65 | real+61 | **e21, He21,** e52, n21, **n52,** n41, **n61** | — | BAD |
| 0.5 | 500 | 80 | real+61 | e52, *n31*, n52, **n61,** r5/2, *r3/5* | — | BAD |
| 0.5 | 500 | 95 | real+61 | **n61,** *r3/5* | — | possibly bad |
| 0.5 | 500 | 145 | 61, L7 | **n61?** | | ok? |
| | | | | | | |
| 0.6 | 140 | 140 | 61, L7 | **n61** | **e51,** *n51*, **r5/3** | possibly bad |
| | | | | | | |
| 1.0 | 175 | 65 | real, L7 | **He21, n21**, **n52, n31, n41, r5/2** | **e21, He21, e52, He52**, *e31*, **e51**, **n21, n52, n31, n51**, r5/2, **r3/5, r5/3** | BAD |
| 1.0 | 175 | 80 | real, L7 | **e52, He52, n52, n31, n41, r5/2, r3/5** | **e52, He52**, **e31**, **e51**, **n31, n51**, **r3/5, r5/3** | BAD |
| 1.0 | 175 | 95 | real+61, L7 | n31, **n41, n61,** r3/5 | e52, e31, **e51**, **n51**, r3/5, **r5/3** | BAD |
| 1.0 | 250 | 65 | real, L7 | **e21, He21,** e52, He52, **n21**, **n52, n31, n41**, **r5/2** | **e21, He21, e52, He52, e31, e51, n21, n52, n31, n51, r5/2,** r3/5, **r5/3** | BAD |
| 1.0 | 250 | 80 | real, L7 | **e52, He52, n52, n31,** n41, **r5/2,** r3/5 | **e52, He52, e31, e51, n31, n51**, r5/2, **r3/5, r5/3** | BAD |
| 1.0 | 250 | 95 | real+61, L7 | **n41**, **n61,** r5/2, *r3/5* | e52, *e31*, e51, **n51**, **r5/3** | BAD |
| 1.0 | 250 | 145 | 61, L7 | **n61** | | marginally ok |
| 1.0 | 250 | 195 | 61, L7#51 | **n61?** | | ok? |
| 1.0 | 500 | 65 | real+61 | **e21, He21, e52, He52, n21, n52, n31, n41, n61, r5/2** | — | BAD |
| 1.0 | 500 | 80 | real+61 | **e52, He52, n52,** n31, **n41**, **n61**, **r5/2** | — | BAD |
| 1.0 | 500 | 95 | real+61 | **n41, n61** | — | BAD |
| 1.0 | 500 | 145 | 61, L7 | **n61** | | marginally ok |
| 1.0 | 500 | 195 | 61, L7#51 | | | **OK** |
| | | | | | | |
| 2.0 | 250 | 65 | real, L7 | **e21, He21, n21, n52, n31, n41,** *r5/2* | **e21, He21, e52, He52,** *e31*, **e51, n21, n52, n31, n51, r5/2, r5/3** | BAD |
| 2.0 | 250 | 80 | real, L7 | **e52,** He52, **n52, n31, n41, r5/2,** r3/5 | **e52, He52, e31, e51**, n52, **n31, n51, r5/2, r3/5, r5/3** | BAD |
| 2.0 | 250 | 95 | real, L7 | **n41,** r5/2, **r3/5** | e52, **e31, e51, n51, r5/3** | BAD |
| 2.0 | 250 | 145 | 61, L7 | **n61** | e51,**r5/3** | BAD |
| 2.0 | 250 | 195 | 61, L7#51 | **n61?** | | ok? |
| 2.0 | 500 | 145 | 61, L7 | **n61** | | marginally ok |
| 2.0 | 500 | 195 | 61, L7#51 | | | **OK** |
| | | | | | | |
| 3.0 | 250 | 195 | 61, L7#51 | **n61** | | marginally ok |
| 3.0 | 500 | 145 | 61, L7 | **n61** | **e51,** r5/3 | possibly bad |
| 3.0 | 500 | 195 | 61, L7#51 | **n61?** | | ok? |

**Notes.** $m_P$, $a_P$, and $q_P$ represent the KBP's mass, semimajor axis, and perihelion, respectively. Inclinations $i_P$ = 0 and 40 deg were considered for all KBPs in this investigation. In the fourth column, 'real' refers to 2:1, 5:2, 3:1, 4:1, 5:1, and 6:1 stable resonant populations from observations, 'L7' refers to 2:1, 5:2, 3:1, and 5:1 stable resonant populations from the L7 model ('L7#51' means that only the 5:1 population was considered). Refer to Table A3 for the codes

used in the fifth and sixth columns. The results are shown according to the KBP's initial inclination as indicated by the text style: bold ($i_P$ = 0, 40 deg), plain ($i_P$ = 0 deg), and italic ($i_P$ = 40 deg). The long bar means that no L7 population was tested. See also Section 2.

**Table A3.** Criteria used to demonstrate Gyr-stable resonant objects that were disrupted by the perturbation of a KBP compared to the baseline results of the control model.

| | | Observed TNO populations after 4 Gyr | L7 populations after 1 Gyr | Observed TNO populations after 4 Gyr | L7 populations after 1 Gyr |
|---|---|---|---|---|---|
| Mean motion resonance (MMR) | Code | KBP model: 4 GPs + one resident KBP | KBP model: 4 GPs + one resident KBP | Control model: only 4 GPs | Control model: only 4 GPs |
| 2:1 | n21 | $n < 0.3$ | $n < 0.25$ | $n = 0.7$ | $n = 0.6$ |
| | e21 | $e < 0.2$ | $e < 0.198$ | $e = 0.262$ | $e = 0.208$ |
| | He21 | $f(q < 33$ au$) < 0.1$ | $f(q < 33$ au$) < 0.05$ | $f(q < 33$ au$) = 0.27$ | $f(q < 33$ au$) = 0.15$ |
| 5:2 | n52 | $n < 0.4$ | $n < 0.4$ | $n = 0.9$ | $n = 0.95$ |
| | e52 | $e < 0.36$ | $e < 0.281$ | $e = 0.406$ | $e = 0.296$ |
| | He52 | $f(q < 33$ au$) < 0.2$ | $f(q < 33$ au$) < 0.03$ | $f(q < 33$ au$) = 0.5$ | $f(q < 33$ au$) = 0.075$ |
| 3:1 | n31 | $n < 0.2$ | $n < 0.2$ | $n = 0.8$ | $n = 0.7$ |
| | e31 | — | $e < 0.316$ | — | $e = 0.366$ |
| 4:1 | n41 | $n < 0.2$ | — | $n = 0.7$ | — |
| 5:1 | n51 | — | $n < 0.15$ | — | $n = 0.6$ |
| | e51 | — | $e < 0.499$ | — | $e = 0.549$ |
| 6:1 | n61 | nominal or > 4 clones not in MMR | — | nominal+8 clones in MMR | — |
| | r5/2 | $r$ (5:2/2:1) < 0.8 | $r$ (5:2/2:1) < 5 | $r$ (5:2/2:1) = 0.98 | $r$ (5:2/2:1) = 7 |
| | r3/5 | $r$ (5:2/2:1) < 0.1 | $r$ (5:2/2:1) < 0.09 | $r$ (3:1/5:2) = 0.23 | $r$ (3:1/5:2) = 0.15 |
| | r5/3 | — | $r$ (5:1/3:1) < 0.45 | — | $r$ (5:1/3:1) = 0.7 |

**Notes.** The control model considered only the four giant planets with their current orbits, providing the baseline results that allowed us to evaluate the effects of a resident KBP on stable resonant populations. The KBP model considered the giant planets and an additional planet with various orbits and masses, as shown in Table A2. The criteria for assessing the perturbation applied by a KBP on these resonant populations are summarized in the third and fourth

columns. *n* indicates the ratio of the resulting population compared to the initial one, *e* is the median eccentricity over a resonant population, *f* is the fraction of objects with perihelion $q < 33$ au in a given resonant population, and *r* represents the ratio of populations for two given MMRs n:m and p:q. Given small-number statistics, the single 6:1 MMR TNO was represented by a nominal member using the best-fit orbit and eight clones covering the orbital uncertainties of that orbit. A missing value indicates that the referred resonant population was unavailable.